\documentclass{aa}
\usepackage{times}
\usepackage{psfig}

\begin{document}
\title{The structure of contact binaries}
\author{H. K\"ahler}
\institute{Hamburger Sternwarte, Gojenbergsweg 112, D-21029 Hamburg, Germany
\,(e-mail: hkaehler@hs.uni-hamburg.de)}
\date{Received  / Accepted}

\abstract{
In radiative layers of rotating stars the luminosity carried by
circulation currents through a surface of constant entropy (circulation luminosity)
is shown to be positive.
The corresponding decrease in the temperature gradient is important in the secondary
of contact binaries. This result removes the deadlock in the theory of
contact binaries.

The resulting treatment of contact binaries is investigated, assuming thermal
equilibrium. The sources of the circulation luminosity in the secondary
can be written as the product of a circulation function (a normalized non-negative
function of the fractional mass) and an amplitude. If the amplitude is adjusted
to give a prescribed temperature difference $\Delta T_{\rm e}=T_{{\rm e}1}-T_{{\rm e}2}$,
the choice of the circulation function is (in a broad range) unimportant.
This invariance extends in a close approximation to all observable properties
as well as to the internal structure.
The temperature difference is bound to be positive. The fractional extent of
radiative regions is larger in the secondary than in the primary.
In the course of evolution the period increases and the mass ratio decreases.
Comparing thermodynamic quantites on level surfaces, pressure and density are
larger in the secondary than in the primary. The specific entropy is larger in the
primary than in the secondary. The temperature difference is remarkably small and
almost vanishing when averaged over the level surfaces occupied in both components.

The only free parameter (apart from $\Delta T_{\rm e}$) is the efficiency $f_{\rm E}$
of the energy transfer from the primary to the secondary.
Using standard values for the parameters,
a survey of unevolved and evolved contact configurations is presented.
Observational tests are passed. In stable systems the degree of contact is small. 
Stable systems in the period-colour diagram, unevolved and evolved, cover the strip 
(and only the strip) of observed systems in this diagram. Lower limits for period and
effective temperature, compatible with the observed limits, are caused
by the requirement of thermal stability. Stable systems with mass ratios very close
to unity are possible, in accordance with recent observations.
Since stability considerations are essential in these observational tests
the results support the assumption of thermal equilibrium as well as
the treatment of the stability problem.

Models for individual observed systems with reliable data are well-determined (apart
from some freedom in $\Delta T_{\rm e}$) and can be used  to calibrate the efficiency
and to determine metallicity and age. All models obtained so far are stable.
This again supports the assumption of thermal stability. The results show that evolutionary effects
are important and that the efficiency is very small ($f_{\rm E}=10^{-3}\ldots 10^{-5}$).

Arguments are presented that the velocity field in the common envelope
has a reversing layer, with motions from the secondary to the primary in the layers
just above the critical surface and from the primary to the secondary in the
surface layers.

}

\maketitle

\section{Introduction}        \label{intro}

This paper is the third one in a systematic attempt to overcome the
deadlock in the theoretial treatment of contact binaries. In the first paper
(K\"ahler \cite{k02a}, hereafter K1) we
discussed the uncertainties in the structure equations. Serious uncertainties
were found to
concern only the energy sources and sinks associated with internal mass motions.
In the second paper (K\"ahler \cite{k02b}, hereafter K2) we made the usual
assumption (hereafter assumption A) that sources and sinks occur only in the common
envelope. Investigating a typical system we found a variety of solutions which are
reasonable from an observational point of view, but all of them suffer from
theoretical deficiencies. In particular, all solutions with good light curves
(i.e. a small temperature
difference between the components) are of the type proposed by Moss \& Whelan (\cite{mwh73}),
i.e. the energy sources in the secondary's envelope occur in the extreme
superadiabatic layers. As shown by Hazlehurst (\cite{ha74}) this is impossible
since the heat capacity of these layers is too small.

The results in K2 show that assumption A is too restrictive.
When thermal equilibrium is assumed this conclusion is cogent since it
does not depend on details, e.g. the choice of a
transport equation for the energy transfer from the primary to the secondary.
In the case of thermal cycles there is more freedom since they depend
on the transport equation as well as on the equation governing the
rate of mass exchange. Nevertheless the experience of decades
shows that the conclusion is justified also in this case.
Associated to a cycle is a configuration
in unstable thermal equilibrium. If this configuration is in shallow contact
(and if extreme superadiabatic transfer is excluded)
the temperature difference is large for a considerable part of the cycle.
If the configuration is in good contact
the temperature difference is always large. In both cases the light curve is bad.

Assumption A turns out to be a deadlock in the treatment of contact binaries. Energy 
sources or sinks in the interior of at least one component and
corresponding changes in the temperature gradient are necessary.
According to Lucy (\cite{lucy68}) there is no freedom in the structure of radiative
regions since the luminosity carried by circulation
currents through a level surface vanishes (hereafter Lucy's theorem).
The only remaining freedom concerns the extent of turbulent regions. A decrease of a convective envelope
is in conflict with Lucy's theorem. An increase of the turbulent envelope
in the secondary is in conflict with Roche geometry. An increase in the primary
is compatible with Roche geometry, but the resulting model
(K\"ahler \cite{k89}, K\"ahler \& Fehlberg \cite{kf91})
is in conflict with the second law of thermodynamics (Hazlehurst \cite{ha93})
since (in a large region) internal mass motions transfer energy from cooler to hotter layers.

Accordingly, the deadlock boils down to Lucy's theorem. This theorem has been
derived --- also by Mestel (\cite{mes}) and Roxburgh et al. (\cite{rox}) ---
assuming that all thermodynamic quantities are constant on equipotential
surfaces, i.e. assuming pseudo-barotropic layers in strict hydrostatic equilibrium.
Although the theorem is usually taken for granted (e.g. Maeder \& Zahn \cite{maeza}),
it is valid only in the zero-order approximation and not in
higher approximations (Mestel \cite{mes}). In contact binaries
departures from strict hydrostatic equilibrium (and thus baroclinic effects) are
necessary for the energy transfer from the primary to the secondary,
and baroclinic effects are expected to be important also in the interior of
the components since the variation of the gravity on a level surface is large.
For these reasons it is likely that Lucy's theorem does not apply.
As observed also by Tassoul \& Tassoul (\cite{tass95}) in a discussion of rotating
stars there is no reason to claim that the net flux of energy carried by the
circulation currents through each level surface vanishes.

Removing the deadlock and assuming thermal equilibrium we are able to
predict the modification of the original Lucy (\cite{lucy68}) model which
solves the contact binary problem.
Roche geometry requires that (in comparison with the Lucy model) either the (averaged) temperature gradient in the primary be increased or that the gradient in the secondary be decreased. An enhanced gradient is in conflict with the second law. Accordingly we need a reduced gradient in the
secondary and thus a reduced radiative luminosity and/or a reduced extent
of convective layers. The reduction is possible
if some part of the luminosity in the secondary's interior is
carried by circulation currents to the common envelope.

In this paper we investigate the resulting treatment of contact binaries, assuming that
thermal equilibrium is enabled by circulation currents in the secondary's interior.
We discuss the effects of circulation, present a survey of unevolved and evolved
theoretical configurations, and perform observational tests. Having obtained
models for individual observed systems we shall finally
return to the hydrodynamic problem.

\begin{figure}
\psscalefirst
\psfig{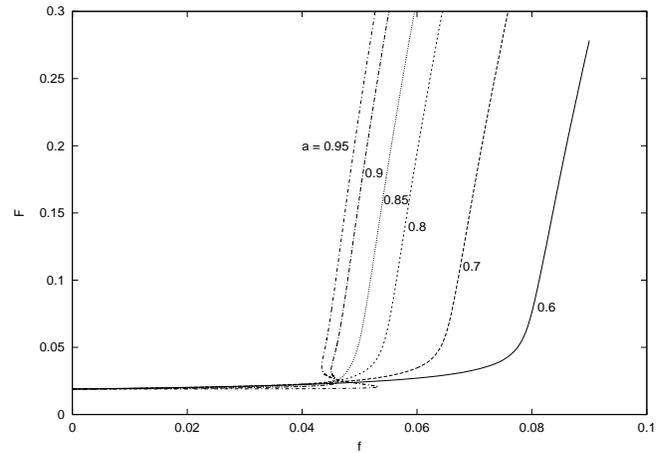}
\caption[]{The degree of contact $F$ in dependence on the  amplitude $f$
of the circulation in System~1,
for different values of the parameter $a$ (see text).}
\label{fig1}
\end{figure}

\begin{figure}
\psscalefirst
\psfig{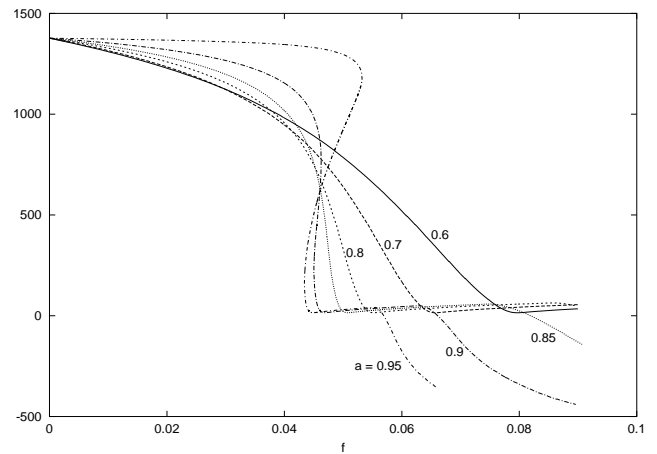}
\caption[]{The temperature difference $\Delta T_{\rm e}$ in dependence
on the  amplitude $f$ in System~1,
for different values of the parameter $a$.}
\label{fig2}
\end{figure}

\begin{figure}
\psscalefirst
\psfig{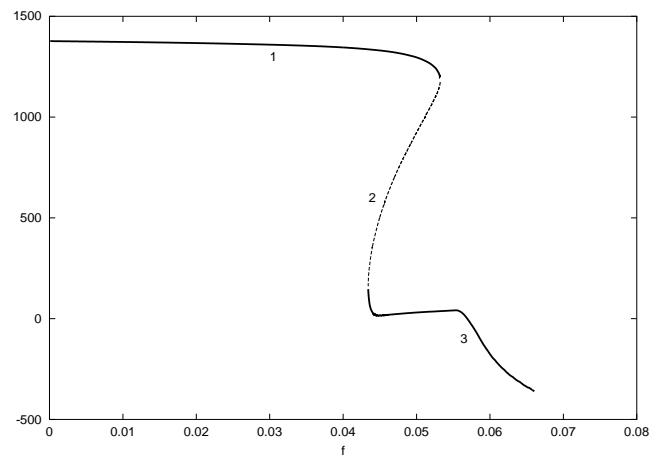}
\caption[]{$\Delta T_{\rm e}$ as a function of $f$ for $a=0.95$ in System~1.
Branches~1 and 3 (solid lines) of the linear series are
stable, while branch~2 (dashed line) is unstable (see text).
}
\label{fig3}
\end{figure}

%%%%%%%%%%%%%%%%%%%%%%%%%%%%%%%%%%%%%%%%%%%%%%%%%%%%%%%%%%%%%%%%%%%%

\section{The effects of circulation currents}                     \label{eff}

We shall adhere to the notation of K1 and K2.
Roche geometry is valid in a close
approximation and will be assumed throughout this paper, apart
from  the discussion of the velocity field of the internal mass motions in
Sect.~\ref{com}.
An equation for the transfer of mass between the components will not
be needed. The energy transfer from the primary to the secondary will be treated as
proposed in K1, assuming $\alpha_1=\alpha_2$.
Recall that $\alpha$ denotes the extent of the energy sinks/sources in the
primary's/secondary's envelope, expressed as fractional mass of the layers
above the critical surface. Since extreme superadiabatic transfer is not possible,
the case of $\alpha\ll 1$ is excluded. Since the transfer occurs
in the common envelope, values of $\alpha$ much larger than unity are also
excluded. The parameter $f_{\rm E}$, measuring the efficiency of the energy transfer,
 is certainly smaller than unity and possibly much smaller.
The mixing length in units of the pressure scale height will taken
to be $\alpha_+=1.5$. We shall assume that the extent of turbulent regions
is determined by the Schwarzschild criterion ($i_t=0$).

With these assumptions we have so far only two free parameters
$\alpha,f_{\rm E}$. An additional parameter will be required when introducing circulation
currents in the secondary. A modification of the structure equations
derived in K1 will also be required.

%%%%%%%%%%%%%%%%%%%%%%%%%%%%%%%%%%%%%%%%%%%%%%%%%%%%%%%%%%%%%%%%%%%%%%%%%%%%%%%%

\subsection{Energy balance and temperature gradient}    \label{energ}

In the common envelope the luminosity $\Lambda$ is transferred from the
primary to the secondary (hereafter process~1). Let us assume that in addition
in the secondary a fraction $f$ of $\Lambda$ is transferred by circulation currents
from the interior to the envelope (process~2). Here we collect the equations
describing the superposition of these processes.

The local balance of energy in the component $i$ is
\begin{equation}                          \label{dg3}
\partial l_i/\partial m_i=
                          \varepsilon_i-T_i\dot{s}_i+\sigma_i,
\end{equation}
where $\sigma_i$ represents the sources or sinks in the envelope caused
by process~1, i.e.by the energy transfer form the primary to the secondary
in a region described by the parameter $\alpha$. (Although thermal equilibrium
is assumed, the time-dependent form of the energy balance is needed when
discussing thermal stability.) Note that $\sigma_2$ does not allow for process~2.

In the presence of circulation in a
radiative region of the secondary the luminosity is the sum
\begin{equation}    \label{sum}
l_2=l_{{\rm rad,}2}+l_{{\rm circ,}2}
\end{equation}
of the radiative luminosity  $l_{{\rm rad,}2}$ and
the luminosity carried by circulation currents $l_{{\rm circ,}2}$
(hereafter circulation luminosity).
The temperature gradient in a radiative layer
\begin{equation}          \label{nab}
\nabla_{{\rm rad,}2}=\frac{3}{16\pi {\rm acG}}
\frac{\kappa_2 l_{{\rm rad,}2}P_2}{m_2 T_2^4}
\end{equation}
is coupled with $l_{{\rm rad,}2}$ and not with $l_2$. (Thus,
in the presence of circulation Eq.~(5) in K1 is not valid.) The borders of
convective regions are therefore also coupled with $l_{{\rm rad,}2}$.
Circulation occurs as well in convective regions (e.g. Kippenhahn \cite{kip63}).

To investigate the effects of process~2 we shall assume a run of
$l_{{\rm circ,}2}$ throughout the secondary as a non-negative function of the
mass variable which vanishes at the boundaries. (Since process~2 amounts to
a redistribution of energy, the sources of the circulation luminosity, i.e. the
regions with a positive derivative ${\rm d}l_{{\rm circ,}2}/{\rm d}m_2>0$,
are compensated by sinks with a negative derivative.)
Equations~(\ref{sum}) and (\ref{nab}) define the radiative gradient
$\nabla_{{\rm rad,}2}$. If this gradient is not larger than the adiabatic gradient
 the layer is radiative. Otherwise the layer is convective and the
temperature gradient can be obtained from the mixing-length theory.
Accordingly, the temperature gradient is determined as usual,
replacing simply the luminosity $l_2$ by the radiative luminosity
$l_{{\rm rad,}2}$. We have tacitly assumed that
 $l_{{\rm rad,}2}$ is positive apart from the centre. Otherwise the
choice of $l_{{\rm circ,}2}$ is unphysical.

Writing the circulation luminosity as a function of the fractional mass
$x_2=m_2/M_2$, we shall assume that the derivative is a superposition
\begin{equation}         \label{superpos}
{\rm d}l_{{\rm circ,}2}/{{\rm d}x_2}=f\Lambda\,c(x_2)+\ldots
\end{equation}
of sources and sinks. In the first term, representing the sources,
$c(x_2)$ is a normalized non-negative function
which will be called circulation function. The amplitude~$f$ is
expected to be small compared to unity in realistic models with typical mass ratios.
The dots represent corresponding sinks in the envelope with a distribution
described again by the parameter $\alpha$.

%%%%%%%%%%%%%%%%%%%%%%%%%%%%%%%%%%%%%%%%%%%%%%%%%%%%%%%%%%%%%%%%%%%%%%%%%%%%%%%%

\subsection{The choice of the circulation function}                \label{inv}
Since so far nothing is known about realistic circulation functions,
two simple functions depending on a parameter $a$ will be used.
Writing for simplicity $x$ instead of $x_2$, the first function
\begin{equation}                            \label{class1}
c_1=\left\{ \begin{array}{ll}
 \left\{1+\cos\left[\pi(x-a)/d \right]\right\}/(2d)&
\quad\mbox{if \, $x>a-d$}\\
 0                                             & \quad \mbox{otherwise} \\
              \end{array}
     \right.
\end{equation}
with $d=x_{\rm f}-a$ (where $x_{\rm f}$ represents the fitting mass)
has its maximum at $x=a$.
The second function
\begin{equation}                            \label{class2}
c_2=\left\{ \begin{array}{ll}
 0                                                      & \quad \mbox{if $x\le a$}     \\
 V\left\{1-\cos\left[\pi (x-a)/(b-a)\right]\right\} & \quad \mbox{if $a<x<b$}          \\
 2V                                                     & \quad \mbox{if $x\ge b$}     \\
              \end{array}
     \right.
\end{equation}
with
\begin{equation}                                     \label{vab}
b=(a+x_{\rm f})/2,\qquad    V=1/(2x_{\rm f}-a-b)
\end{equation}
has a non-negative derivative, vanishes at the centre, and increases between
$x=a$ and $x=b$.

\begin{figure}
\psscalefirst
\psfig{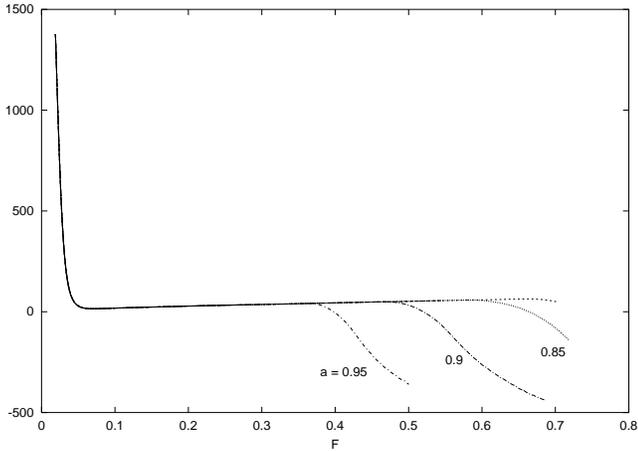}
\caption[]{The temperature difference $\Delta T_{\rm e}$ as a function of $F$ in System~1,
for the same values of the parameter $a$ as in Figs.~\ref{fig1} and \ref{fig2}.
}
\label{fig4}
\end{figure}

\begin{figure}
\psscalefirst
\psfig{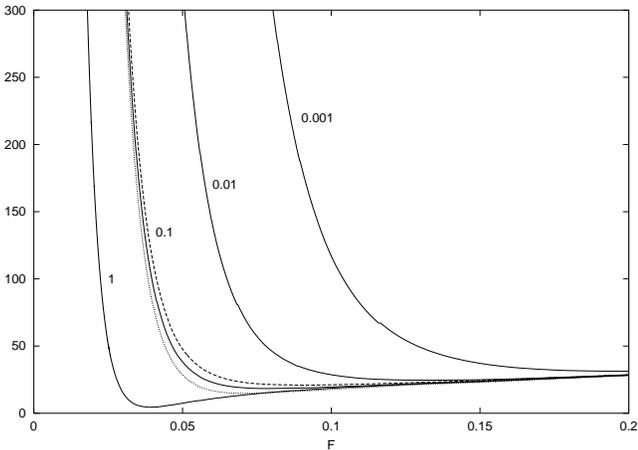}
\caption[]{$\Delta T_{\rm e}$ as a function of $F$ in System~1,
for  $\alpha=1$ and different values of $f_{\rm E}$ (solid lines),
for $\alpha=0.5,f_{\rm E}=0.1$ (dotted line),
and for $\alpha=5,f_{\rm E}=0.1$ (dashed line).
}
\label{fig5}
\end{figure}

As in K2 we investigate an unevolved system with the physical parameters
\begin{equation}
M=1.6 M_{\sun},\; J_{52}=0.4,\; X=0.7,\; Z=0.02,
\end{equation}
where $J_{52}$ denotes the angular momentum in $10^{52}$ cgs-units
(hereafter system~1).
Using the parameters $\alpha=0.5,f_{\rm E}=0.1$ and the
circulation functions $c_1(a,x)$ with different values for $a$ we
calculated a continuous sequence of configurations 
in dependence on the amplitude $f$ of the circulation, i.e. a linear series
with the parameter $f$.
Results for the temperature difference  $\Delta T_{\rm e}=T_{{\rm e}1}-T_{{\rm e}2}$
and the degree of contact $F$  are shown in Figs.~\ref{fig1} and~\ref{fig2}.
In the absence of circulation ($f=0$) the temperature difference
is very large and the contact is shallow. We need configurations in shallow
contact with a small temperature difference. For any choice of $a$ they can be
obtained with a small amplitude ($f\simeq 0.04\ldots 0.08$).
This is as expected from the arguments collected in Sect.~\ref{intro}.
Unfortunately the curves depend sensitively on $a$, i.e. on the choice of
the circulation function which so far is unknown.

In particular, note that the linear series has
turning points if $a$ is sufficiently large ($a\ge 0.9$), i.e. if sources of
the circulation luminosity occur only in outer layers. For $a=0.95$ this is
illustrated in Fig.~\ref{fig3}. The curve consists of three branches.
Configurations on branch~2 (dashed line) are unstable.
Configurations which are reasonable from
an observational viewpoint (in the neighbourhood of the turning point
between branches 2 and 3) are either unstable or close to instability.
Unstable configurations are unrealistic. Solutions which are stable but
close to instability are extremely fragile from a computational viewpoint
and thus probably also unrealistic. This suggests that the sources of the circulation
luminosity occur not only in outer layers but also deep in the secondary's interior.

The uncertainty in the choice of the circulation function
seems to be a serious drawback. However, if $f$ is eliminated and  $\Delta T_{\rm e}$
considered as a function of $F$,
all curves coincide apart from very large values for $F$ which are not of interest.
This surprising result is shown in Fig.~\ref{fig4}. In other words, the relation
between the two
observable quantities  $\Delta T_{\rm e}$ and $F$ does not depend on the
choice of the circulation function.
This invariance (which has been checked in numerous cases
and which extends much further as we shall see) enables a treatment of contact
binaries although so far we are not able to determine details of the circulation.

Figure~\ref{fig5}
shows the relation between $\Delta T_{\rm e}$ and $F$ for system~1
in the region of interest (a small temperature difference and
shallow contact) in more detail and for different values of the parameters
$\alpha$ and $f_{\rm E}$. The dependence on $\alpha$ is weak and can be
neglected in an approximate treatment, but the dependence on $f_{\rm E}$ is
important. The curve for $f_{\rm E}=1$ represents an unrealistic limiting case.
Realistic values for $f_{\rm E}$ are certainly smaller than unity and
probably much smaller. The curves show therefore that  the
temperature difference in system~1 is positive.

%%%%%%%%%%%%%%%%%%%%%%%%%%%%%%%%%%%%%%%%%%%%%%%%%%%%%%%%%%%%%%%%%%%%%%%%%%%%%%%%%%%%%%%

\begin{figure}
\psscalefirst
\psfig{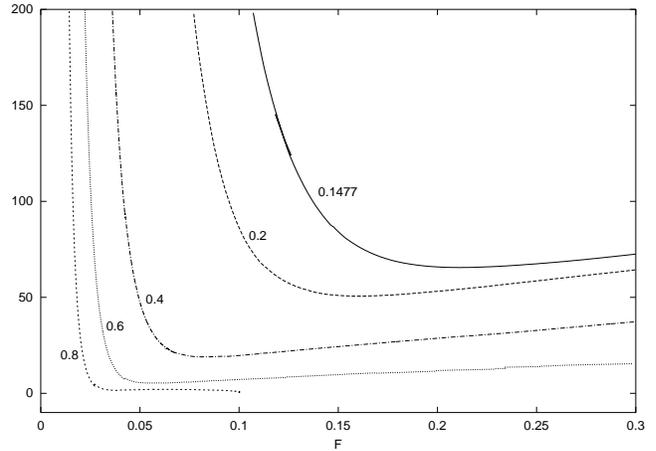}
\caption[]{Lines of constant angular momentum in the $(\Delta T_{\rm e},F)$ diagram
for systems with $M=1.6 M_{\sun}$. The numbers denote the mass ratio for
$\Delta T_{\rm e}=200$~K (see text).
}
\label{fig6}
\end{figure}

\subsection{The temperature difference}              \label{tdif}
Having seen that the temperature difference in system~1 is bound to be positive,
here we ask whether this result can be generalized. As an example we consider unevolved
systems of the same mass and composition as before, with $\alpha=1, f_{\rm E}=0.1$.
Figure~\ref{fig6} shows the invariant relation
between $\Delta T_{\rm e}$ and $F$ in systems with  different
angular momentum. The curves are labeled by the mass ratio for $\Delta T_{\rm e}=200$~K.
The curve beginning with $\Delta T_{\rm e}=200$~K and $q=0.8$ ends up with
$\Delta T_{\rm e}\rightarrow 0$ and   $q\rightarrow 1$, i.e. with
a system with equal components.

It is manifest that the temperature difference is positive
apart from the limiting case $q\rightarrow 1$. This result has been checked in many
other cases.
It is valid for all stable systems, evolved or unevolved.

Concering a physically plausible interpretation, circulation currents
tend to decrease the departures from hydrostatic equilibrium in the whole system. Therefore
they tend to reduce the temperature difference, which means that
\begin{equation}                              \label{tend}
{\rm d}\Delta T_{\rm e}/{\rm d}f<0,
\end{equation}
as it is the case in the relevant parts of the curves in Fig.~\ref{fig2} unless
the branch is unstable, i.e. unless the parameter $a$ is too large.
Since the circulation currents do not quite succeed in removing the temperature
difference, this difference remains positive.

In the example just considered the system with equal components is unstable, 
but systems with mass ratios up
to $q=0.9918$ are stable. For a very long time the absence of systems
with nearly equal components seemed to be an established observational fact,
and an instability of systems with $q\simeq 1$ was discussed as a possible
theoretical explanation. The present example shows that the existence of systems with
mass ratios very close to unity cannot be excluded from a theoretical viewpoint. This is
in accordance with the observation of the system V 753 Mon with $q\simeq 0.97$
(Rucinski et al. \cite{ru00}).

%%%%%%%%%%%%%%%%%%%%%%%%%%%%%%%%%%%%%%%%%%%%%%%%%%%%%%%%%%%%%%%%%%%%%%%%%%%%%%%%%%%%%%%%%

\begin{table*}
\caption[ ]{Observable properties of system~1\, for
$\alpha=1,f_{\rm E}=0.1,\Delta T_{\rm e}=200$ K  and different
circulation functions (see text).} \label{obs}
\begin{flushleft}
\begin{tabular}{lllllllllll}
\hline
\noalign{\smallskip}
 & $a$ & $P$ & $q$ & $T_{\rm e}$ & $L$ & $L_1$ & $L_2$ & $R_1$ & $R_2$ & $F$   \\
\noalign{\smallskip}
\hline
\noalign{\smallskip}
$c_1$ & 0.5 & 0.2895 & 0.4195 & 5465 & 1.125 & 0.812 & 0.313 & 0.983 & 0.657 & 0.0341 \\
$c_1$ & 0.6 & 0.2890 & 0.4196 & 5462 & 1.120 & 0.808 & 0.312 & 0.982 & 0.656 & 0.0342 \\
$c_1$ & 0.7 & 0.2886 & 0.4195 & 5460 & 1.116 & 0.805 & 0.311 & 0.981 & 0.656 & 0.0342 \\
$c_1$ & 0.8 & 0.2883 & 0.4196 & 5459 & 1.114 & 0.804 & 0.310 & 0.980 & 0.655 & 0.0343 \\
$c_2$ & 0   & 0.2889 & 0.4194 & 5462 & 1.119 & 0.808 & 0.311 & 0.982 & 0.656 & 0.0342 \\
$c_2$ & 0.1 & 0.2888 & 0.4193 & 5461 & 1.118 & 0.807 & 0.311 & 0.981 & 0.656 & 0.0343 \\
$c_2$ & 0.2 & 0.2886 & 0.4193 & 5461 & 1.117 & 0.806 & 0.311 & 0.981 & 0.656 & 0.0343 \\
$c_2$ & 0.3 & 0.2885 & 0.4192 & 5460 & 1.116 & 0.805 & 0.310 & 0.981 & 0.656 & 0.0343 \\
\noalign{\smallskip}
\hline
\end{tabular}
\end{flushleft}
\end{table*}

\begin{table*}
\caption[ ]{Observable properties of system~2\, for
$\alpha=1,f_{\rm E}=0.1,\Delta T_{\rm e}=200$ K  and different
circulation functions.} \label{obs2}
\begin{flushleft}
\begin{tabular}{lllllllllll}
\hline
\noalign{\smallskip}
 & $a$ & $P$ & $q$ & $T_{\rm e}$ & $L$ & $L_1$ & $L_2$ & $R_1$ & $R_2$ & $F$   \\
\noalign{\smallskip}
\hline
\noalign{\smallskip}
$c_1$ & 0.5 & 0.4182 & 0.3606 & 6368 & 4.024 & 2.994 & 1.030 & 1.397 & 0.873 & 0.0599 \\
$c_1$ & 0.6 & 0.4179 & 0.3606 & 6367 & 4.018 & 2.989 & 1.029 & 1.396 & 0.873 & 0.0600 \\
$c_1$ & 0.7 & 0.4177 & 0.3605 & 6366 & 4.014 & 2.987 & 1.028 & 1.396 & 0.872 & 0.0600 \\
$c_1$ & 0.8 & 0.4176 & 0.3603 & 6366 & 4.012 & 2.985 & 1.027 & 1.396 & 0.872 & 0.0601 \\
$c_2$ & 0   & 0.4177 & 0.3606 & 6367 & 4.017 & 2.989 & 1.029 & 1.396 & 0.872 & 0.0599 \\
$c_2$ & 0.2 & 0.4177 & 0.3605 & 6367 & 4.015 & 2.987 & 1.028 & 1.396 & 0.872 & 0.0599 \\
$c_2$ & 0.4 & 0.4175 & 0.3604 & 6367 & 4.013 & 2.985 & 1.027 & 1.395 & 0.872 & 0.0600 \\
$c_2$ & 0.6 & 0.4175 & 0.3603 & 6366 & 4.012 & 2.985 & 1.026 & 1.395 & 0.872 & 0.0601 \\ 
\noalign{\smallskip}
\hline
\end{tabular}
\end{flushleft}
\end{table*}

\begin{table*}
\caption[ ]{Observable properties of system~1 in dependence
on the free parameters.} \label{obs1}
\begin{flushleft}
\begin{tabular}{llllllllllll}
\hline
\noalign{\smallskip}
 $\alpha$ & $f_{\rm E}$ & $\Delta T_{\rm e}$ & $P$ &
$q$ & $T_{\rm e}$ & $L$ & $L_1$ & $L_2$ & $R_1$ & $R_2$ & $F$  \\
\noalign{\smallskip}
\hline
\noalign{\smallskip}
0.5 & 0.1  & 200 & 0.290 & 0.419 & 5461 & 1.122 & 0.810 & 0.312 & 0.984 & 0.657 & 0.033 \\
1   & 0.1  & 200 & 0.289 & 0.419 & 5462 & 1.119 & 0.808 & 0.311 & 0.982 & 0.656 & 0.034 \\
2   & 0.1  & 200 & 0.288 & 0.420 & 5463 & 1.118 & 0.806 & 0.311 & 0.980 & 0.656 & 0.035 \vspace{0.1cm} \\

1   & 0.0001 & 200 & 0.280 & 0.428 & 5435 & 1.086 & 0.776 & 0.309 & 0.971 & 0.660 & 0.141 \\
1   & 0.001  & 200 & 0.284 & 0.419 & 5450 & 1.101 & 0.791 & 0.310 & 0.975 & 0.658 & 0.088 \\
1   & 0.01   & 200 & 0.287 & 0.419 & 5458 & 1.111 & 0.800 & 0.311 & 0.978 & 0.656 & 0.055 \\
1   & 0.1    & 200 & 0.289 & 0.420 & 5462 & 1.119 & 0.808 & 0.311 & 0.982 & 0.656 & 0.034 \vspace{0.1cm} \\
1   & 0.1    & 100 & 0.285 & 0.423 & 5469 & 1.106 & 0.779 & 0.327 & 0.972 & 0.653 & 0.040 \\
1   & 0.1    & 200 & 0.289 & 0.419 & 5462 & 1.119 & 0.808 & 0.311 & 0.982 & 0.656 & 0.034 \\
1   & 0.1    & 300 & 0.292 & 0.416 & 5455 & 1.131 & 0.835 & 0.296 & 0.990 & 0.659 & 0.031 \\
\noalign{\smallskip}
\hline
\end{tabular}
\end{flushleft}
\end{table*}

\begin{figure}
\psscalefirst
\psfig{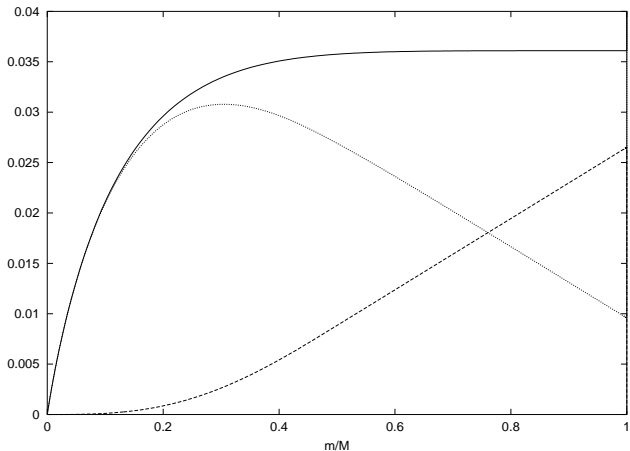}
\caption[]{Luminosities (in solar units) in the secondary's interior in System~2
 in dependence on the fractional mass. The standard circulation function is used.
 The total luminosity (solid line) is the sum of the radiative luminosity (dotted line)
and the circulation luminosity (dashed line).}
\label{fig7}
\end{figure}

\begin{figure}
\psscalefirst
\psfig{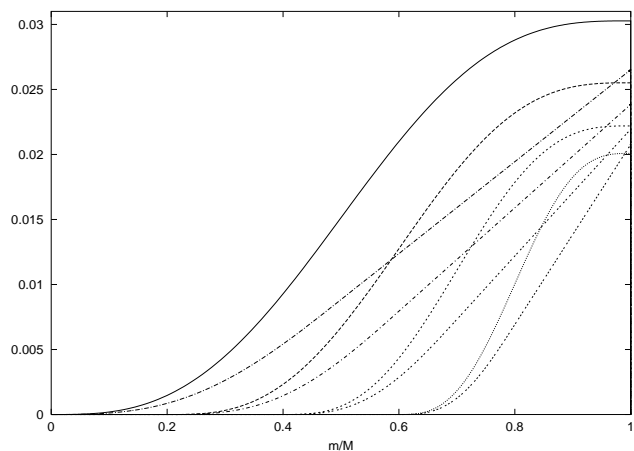}
\caption[]{The circulation luminosity in the interior of the secondary in System~2,
in dependence on the fractional mass,
for different circulation functions (see text).}
\label{fig8}
\end{figure}

\begin{figure}
\psscalefirst
\psfig{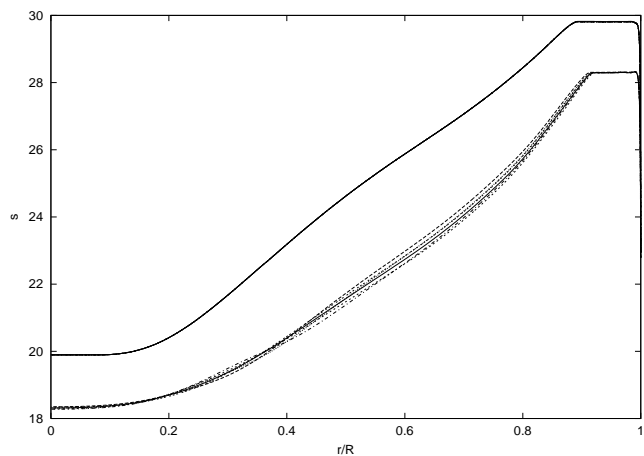}
\caption[]{The entropy distribution in System~2
in dependence on the fractional radius,
for different circulation functions.
The upper curves, representing the primary, coincide. The lower curves
represent the secondary. Flat parts indicate convective regions.}
\label{fig9}
\end{figure}

\begin{figure}
\psscalefirst
\psfig{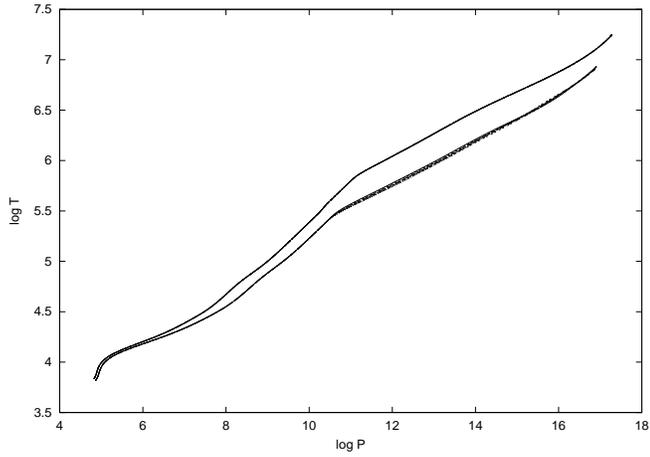}
\caption[]{The $\log P - \log T$ diagram in System~2,
for different circulation functions. $P$ is the pressure in cgs-units.
The upper curves, representing the primary, coincide. The lower curves
represent the secondary.}
\label{fig10}
\end{figure}

\subsection{The treatment of contact binaries}               \label{model}

The discussion in the preceding subsections has shown that in systems with
typical mass ratios a small circulation luminosity (few percent of $\Lambda$) in the
secondary's interior has a large influence on the structure.
A similar circulation luminosity in the primary's interior has almost no effect since
the radiative luminosity is much larger. For this reason circulation in the
primary will be neglected, keeping in mind that in systems with large mass ratios
the situation may be more complex.

To allow for circulation in the secondary we have to specify (i) a circulation
function and (ii) the amplitude $f$ or another measure for the strength of the
circulation. Figure~\ref{fig2} shows that the amplitude which is necessary for a
given temperature difference depends on the choice of the circulation function.
To obtain a measure which is invariant
the parameter $f$ will be replaced by $\Delta T_{\rm e}$. In other words, the amplitude $f$ will be
adjusted to give the precsribed value for $\Delta T_{\rm e}$. The procedure is unique
since in the case of a reasonable circulation function ($a<0.85$ in the example
considered) and a positive temperature difference the connection between
amplitude and temperature difference is unique.

The resulting treatment of contact binaries is essentially based on the assumption
of thermal stability. For given physical parameters (and mixing length) the structure of 
a configuration depends on the choice of the free parameters
$\alpha,f_{\rm E},\Delta T_{\rm e}$
and on the choice of the circulation function. How large are the resulting uncertainties?
We begin using the values
\begin{equation}                 \label{stand}
\alpha=1,\; f_{\rm E}=0.1,\; \Delta T_{\rm e}=200 {\rm K}.
\end{equation}
Keeping the parameters fixed, let us ask whether the observable properties
depend on the choice of the circulation function. Results for system~1 are listed
in Table~\ref{obs}. The first column contains the circulation function, the second one the
parameter $a$. Listed are the period $P$ (in days), the mass ratio $q$, the mean effective
temperature $T_{\rm e}$, the total luminosity $L$ (in solar units), the luminosities and
radii of the components (in solar units), and the degree of contact $F$. 
Table~\ref{obs2}
contains the corresponding results for the system with
\begin{equation}
M=2 M_{\sun},\; J_{52}=0.6,\; X=0.7,\; Z=0.02,
\end{equation}
(hereafter system~2). Larger values for $a$ than those included in the tables lead to
configurations which are either close to instability or unstable. This shows again
that the sources of the circulation extend deep into the secondary's interior.

Tables~\ref{obs} and \ref{obs2} show that the results for different circulation functions
are in close agreement. The function $c_1$ with $a=0.5$ represents a limiting case ($a$
has the smallest possible value).
If the results for this function (first line in the tables)
is omitted, the agreement becomes excellent. We conclude that the choice of the circulation
function (in a broad range) is unimportant as far as observable  quantities are concerned.
In other words, the invariance encountered in Sect.~\ref{inv} extends in a close approximation
to all observable properties.
This invariance suggests the use of a standard circulation function. We decided to use the
function $c_2$ with $a=0$.

Consider next the effects of uncertainties in the free parameters.
Results for system~1, obtained using the standard circulation function,
are listed in Table~\ref{obs1}, which is expected to cover
the range of possible variations of the parameters.
Changes in $\alpha$ have little influence, in accordance with a result from Fig.~\ref{fig5}.
An decrease in the efficiency $f_{\rm E}$ leads mainly to an increase in $F$
(as it might have been expected) and to a decrease in the primary's luminosity.
A change in the temperature difference affects all observable properties, 
particularly the light ratio.

Since these effects are only moderate, an approximate treatment
of contact binaries is possible using standard values of the free parameters.
When getting a first survey of contact configurations in Sect.~\ref{surv}
we shall use the values given in Eq.~(\ref{stand}).
In a next step observational
tests can be used to improve these standard values. In particular, Table~\ref{obs1}
shows that reliable observed values for $F$ can be used to calibrate the efficiency.
This will be done in Sect.~\ref{test}.

%%%%%%%%%%%%%%%%%%%%%%%%%%%%%%%%%%%%%%%%%%%%%%%%%%%%%%%%%%%%%%%%%%%%%%%%%%%%%%%

\begin{figure}
\psscalefirst
\psfig{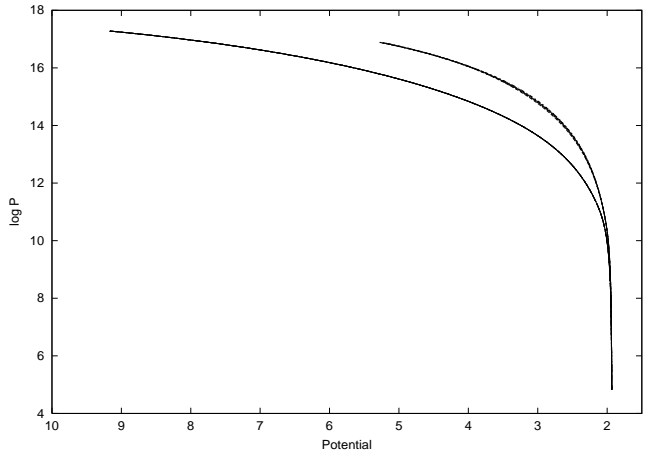}
\caption[]{$\log P$ as a function of the potential in System~2,
for different circulation functions (see text).
The lower curves, representing the primary, coincide. The upper curves
represent the secondary.}
\label{fig11}
\end{figure}

\begin{figure}
\psscalefirst
\psfig{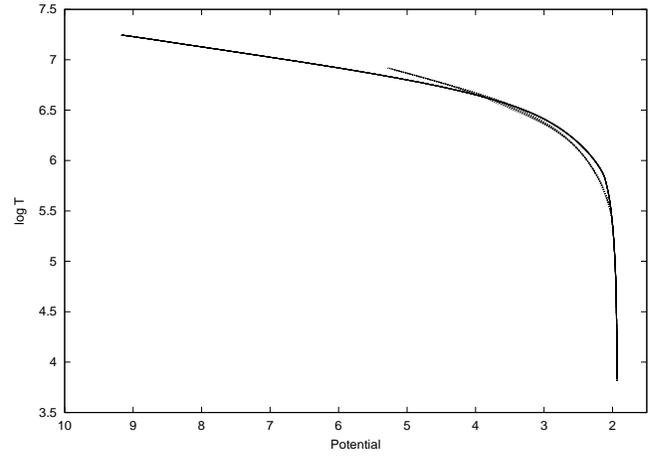}
\caption[]{$\log T$ as a function of the potential in System~2,
for different circulation functions.
The solid curves, representing the primary, coincide. The dotted curves
represent the secondary.}
\label{fig12}
\end{figure}

\begin{figure}
\psscalefirst
\psfig{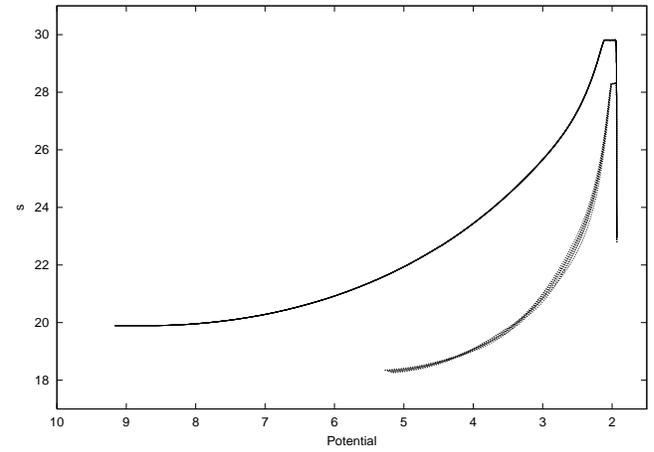}
\caption[]{The specific entropy $s$ as a function of the potential in System~2,
for different circulation functions.
The solid curves, representing the primary, coincide. The dotted curves
represent the secondary.}
\label{fig13}
\end{figure}

\subsection{The internal structure of the components} \label{int}

We shall first consider system~2 as a typical example for the internal structure of
the components and for the influence of the choice of the circulation function.
Figure~\ref{fig7} shows the luminosity in the secondary's interior as
the sum of the radiative luminosity and the circulation luminosity.
(The increase of the luminosity in the envelope escapes notice in this diagram.)
The standard circulation function is used. Figure~\ref{fig8} shows the
circulation luminosity in the secondary's interior for all circulation
functions listed in  Table~\ref{obs2}. It is manifest that two classes
of functions are used.
The curve for the function $c_1$ with $a=0.5$
(solid line) is untypical since the amplitude $f$ is rather large.
As mentioned already this function represents a limiting case which is unlikely
to be realistic. In the further discussion this function will be omitted.
In the following diagrams for system~2
the remaining seven circulation functions are used.

Figure~\ref{fig9} shows the distribution of entropy (per nucleus divided by
Boltzmann's constant). Flat parts of the curves
indicate convective regions. The curves for the primary coincide. 
The curves for the secondary coincide only in the envelope. The uncertainty in 
deeper layers is not large. Note that the fractional extent in radius of the 
convective envelope is larger in the 
primary than in the secondary.

The $\log P - \log T$ diagram (Fig.~\ref{fig10}) illustrates the close correlation
of the two components in the outer layers.
Again the curves for the primary coincide.
The uncertainties in the secondary's curves are remarkably small.

Recall that in both components an effective central potential is well-defined, c.f. K1.
Near the surface this potential coincides with the spherically averaged Roche potential.
Recall furthermore that in the idealized case of strict hydrostatic equilibrium
all thermodynamic quantities are constant on connected
equipotential surfaces. Departures from
strict equilibrium are necessary for the energy transfer. In particular, it is possible
that circulation currents in the common envelope are driven by
pressure differences on equipotential surfaces. Concerning the adjustment of circulation
currents in the interior energy considerations might be important.

\begin{figure}
\psscalefirst
\psfig{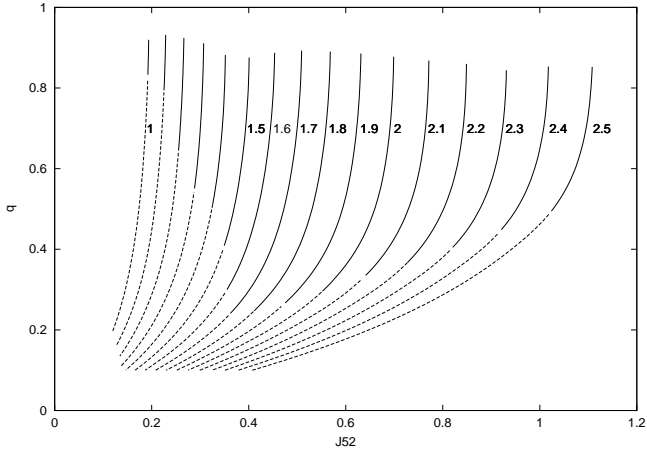}
\caption[]{The mass ratio $q$ in dependence on the angular momentum $J_{52}$
for unevolved systems with $Z=0.02$ (see text).
The numbers denote the mass in solar units. Solid/dahed lines represent
systems of positive/negative charge.}
\label{fig14}
\end{figure}

\begin{figure}
\psscalefirst
\psfig{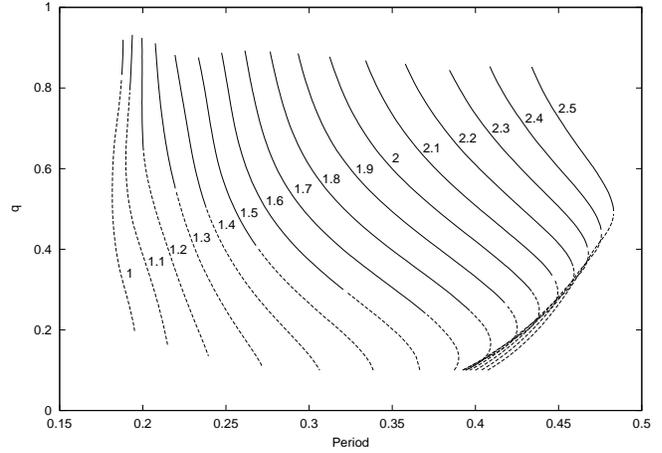}
\caption[]{The period-mass ratio  diagram for unevolved systems with $Z=0.02$.
The symbols are as in Fig.~\ref{fig14}.}
\label{fig15}
\end{figure}

For these reasons it is of interest to investigate thermodynamic quantities in
dependence of the potential. We shall use the normalized potential which is positive,
somewhat smaller than 2 at the surface, and larger in the interior.
Figure~\ref{fig11} shows $\log P$ as a function of the potential. The pressure
difference $\Delta P=P_1-P_2$ is negative throughout the system,
including the outer layers. We found that this result is valid for any system, 
unevolved or evolved, for any mass and any mass ratio.
Implications for the mass motions in the envelope
will be discussed in Sect.~\ref{com}.

Figure~\ref{fig12} shows $\log T$ in dependence on the potential.
The temperature difference $\Delta T=T_1-T_2$ is positive in the outermost layers,
changes sign in deeper layers, and is always small. When averaged over all layers
the difference is very small. This surprising result was again found to be valid for any
system.

Combining the results for $\Delta P$ and  $\Delta T$   and making
use of the equation of state we conclude that the difference in density
$\Delta \rho=\rho_1-\rho_2$ is negative throughout the system.
Consider finally the entropy difference $\Delta s=s_1-s_2$
(Fig.~\ref{fig13}). This difference is positive everywhere and large apart from
the envelope.

From these results concerning differences of thermodynamic quantities on
equipotential surfaces we see that most differences
are large and have a characteristic sign.
An exception is the difference in temperature which is small and almost vanishing when
averaged over the level surfaces occupied in both components.
The temperature is closely connected with the internal energy.
A connection between the total energy of the system  and the thermal adjustment
suggests itself. We shall return to this point in a forthcoming paper.

\begin{figure}
\psscalefirst
\psfig{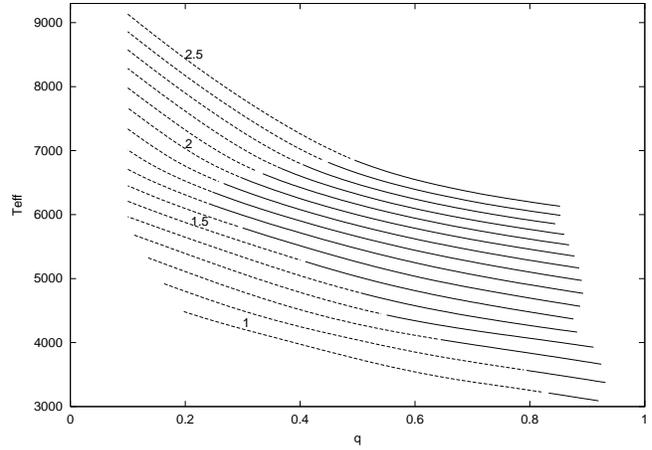}
\caption[]{The mean effective temperature in dependence on the mass ratio $q$
for unevolved systems with $Z=0.02$.}
\label{fig16}
\end{figure}

\begin{figure}
\psscalefirst
\psfig{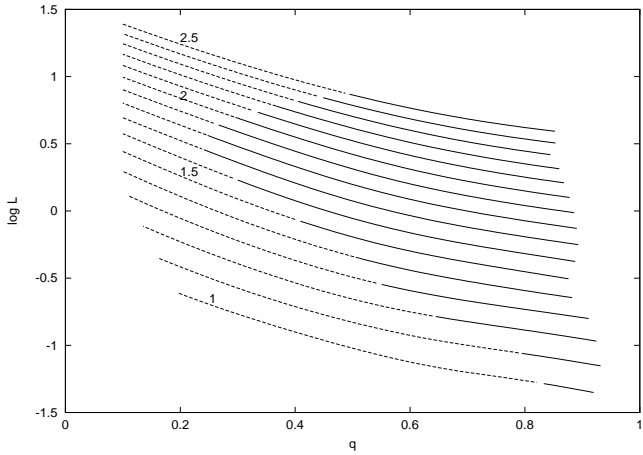}
\caption[]{$\log L$ (where $L$ is the total luminosity in solar units) in dependence 
on the mass ratio $q$
for unevolved systems with $Z=0.02$.}
\label{fig17}
\end{figure}

\begin{figure}
\psscalefirst
\psfig{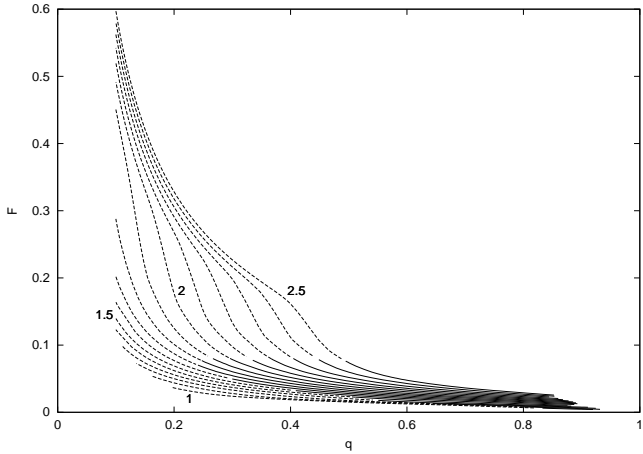}
\caption[]{The degree of contact $F$ in dependence on the mass ratio $q$
for unevolved systems with $Z=0.02$.}
\label{fig18}
\end{figure}

This concludes the discussion of system~2. Results for system~1 are similar,
with slightly larger uncertainties. Both systems have intermediate masses and
mass ratios, and both systems are stable. When passing over to
systems with smaller masses and/or smaller mass ratios, the uncertainties may increase
and the stability may be lost. An example is the system with $M=1 M_{\sun}$ and $q= 0.32$.
Since the secondary's mass is very small
there is a strong tendency towards convection and thus little freedom
for the effects of circulation in radiative regions.
Entropy distribution and number and extent
of convective regions depend strongly on the choice of the circulation function.
Accordingly, the uncertainties in the internal structure are rather large in this
case. These uncertainties do not extend to observable quantities.
They are well defined. Uncertainties in the interior are not of interest
since all configurations are unstable.

Summarizing, in stable systems the invariance found in the preceding
subsections extends also to the internal structure of the components.

%%%%%%%%%%%%%%%%%%%%%%%%%%%%%%%%%%%%%%%%%%%%%%%%%%%%%%%%%%%%%%%%%%%%%%%%%%%%%

\subsection{Remarks on the thermal stability problem}   \label{stab}

Recall first how the thermal stability of a configuration can be tested.
A full test requires either a stability analysis (i.e. the inspection of eigenvalues) 
or evolutionary calculations, starting with small perturbations.
A restricted test requires only the charge $c$ of a configuration, which is the sign of
the Henyey determinant apart from a factor depending on the choice of the Henyey matrix.
The charge $c=1$ is necessary (but not sufficient) for stability, and the charge
$c=-1$ indicates instability.

Consider the second test, i.e. the evolution following a small perturbation.
The evolution can  be calculated if the balance of energy
and the temperature gradient are  known at each point of time.
In other words, we need the sources/sinks $\sigma_i$ and the run of the
circulation luminosity at each point of time.

The sources/sinks $\sigma_i$ are approximately known if the transport equation for
$\Lambda$ is reasonable. Assuming  that the transport equation here used [Eq.~(49) in K1]
is reasonable, this means that the choice of $f_{\rm E}$
is reasonable. The circulation luminosity is known if the circulation function
(represented by the parameter $a$) and the amplitude $f$ are known. Since there is a
large freedom in the choice of the circulation function $a$
can be kept fixed, but changes in $f$ may be important.

\begin{figure}
\psscalefirst
\psfig{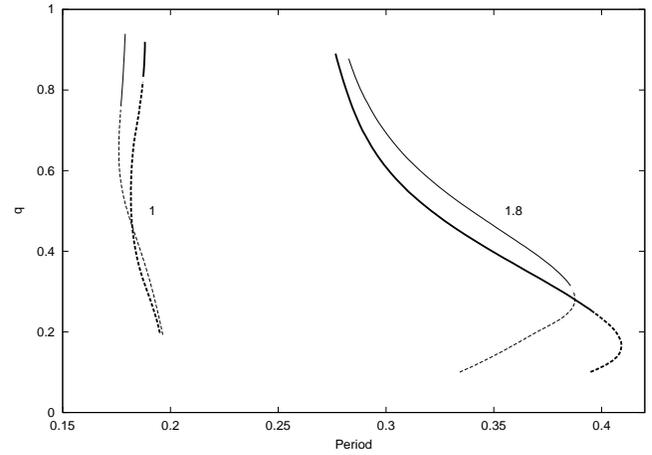}
\caption[]{Unevolved systems with $Z=0.02$ (heavy lines) and $Z=0.01$ (weak lines) in
the period-mass ratio diagram. Solid/dashed lines indicate configurations of
positive/negative charge. The numbers denote the mass in solar units.}
\label{fig19}
\end{figure}

\begin{figure}
\psscalefirst
\psfig{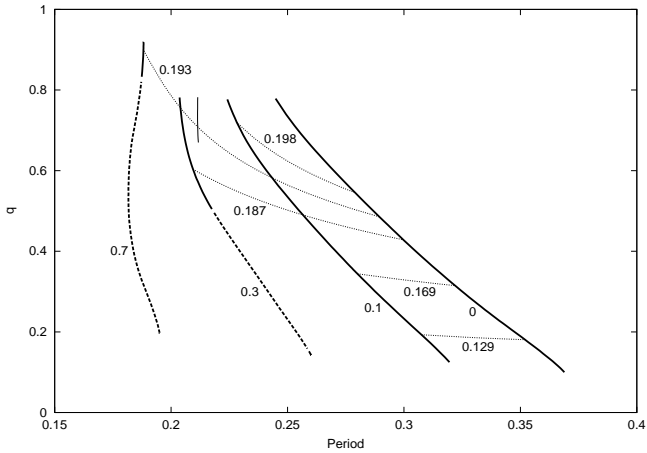}
\caption[]{Evolutionary effects in the period-mass ratio diagram
for systems with $M=1M_{\sun}$ and $Z=0.02$ (see text).
Shown are curves of constant hydrogen content
in the primary's centre (heavy lines) and curves of constant
angular momentum (dotted lines). The numbers denote $X_{{\rm c}1}$ and $J_{52}$, respectively.
Solid and dashed lines are defined as in Fig.~\ref{fig19}.}
\label{fig20}
\end{figure}

\begin{figure}
\psscalefirst
\psfig{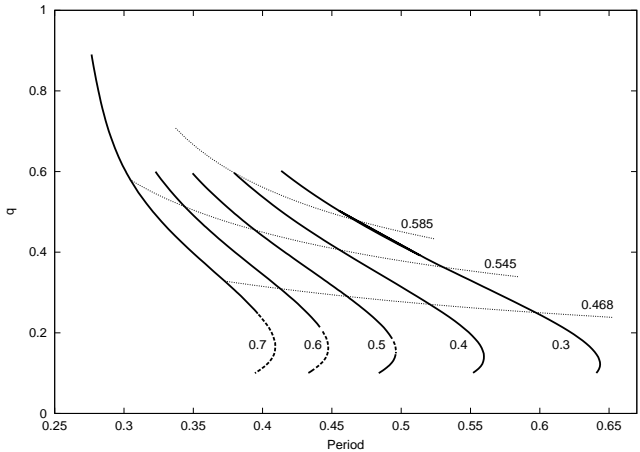}
\caption[]{Same as Fig.~\ref{fig20}, but for systems with $M=1.8M_{\sun}$.}
\label{fig21}
\end{figure}

Throughout this paper we assume that $f$ is kept fixed during a perturbation.
The resulting stability problem is well-defined but simplified.
To make this point clear note that the present definition of the amplitude $f$
is arbitrary and that other definitions are
possible. For example, the expression $f\Lambda$ in the RHS of Eq.~(\ref{superpos})
could be replaced by  $fL$ or by  $fL_{\sun}$ or
by some other expression.
All these possibilities are equivalent as far as the structure of a configuration
in thermal equilibrium is concerned since the amplitude is adjusted to the
prescribed temperature difference. They are however {\em not}
equivalent in the stability problem with a fixed amplitude.
The resulting uncertainty in the stability problem remains to be investigated,
but the results obtained so far suggest that the stability problem here used
is reasonable.

All configurations of positive charge which have been tested so far turned out
to be stable. For simplicity a configuration is therefore called stable if $c=1$.
If the stability has been tested by evolutionary calculations, this is mentioned.

%%%%%%%%%%%%%%%%%%%%%%%%%%%%%%%%%%%%%%%%%%%%%%%%%%%%%%%%%%%%%%%%%%%%%%

\section{A survey of contact configurations}     \label{surv}

In this section we present a survey of contact configurations, adopting
the standard circulation function and the standard values (\ref{stand}) for the
free parameters. Under these circumstances an unevolved configuration
depends on three parameters (mass, angular momentum, metallicity).
An evolved configuration depends in addition on the chemical profile in
both components. In the simplest treatment of systems with intermediate
or small mass ratios a simple chemical profile
is assumed and evolutionary effects in the secondary are neglected.
The evolutionary status is then described by the primary's central hydrogen
content, and we are left with four parameters.

Since a complete survey is impossible we decided to discuss all unevolved
configurations (in a certain mass range) with a standard composition,
and to illustrate evolutionary
effects and the effects of changes in metallicity in few examples. Further examples
are given by the observed systems discussed in Sect.~\ref{test}.
Examples for the effects of changes in the free parameters have already been
discussed.

%%%%%%%%%%%%%%%%%%%%%%%%%%%%%%%%%%%%%%%%%%%%%%%%%%%%%%%%%%%%%%%%%%%%%%%%%%%%%%

\subsection{Unevolved configurations with $Z=0.02$} \label{unev}

We investigate systems with $X=0.07,Z=0.02$ in the mass range
$1\le M/M_{\sun} \le 2.5$. The lower mass limit is caused by the lack of
opacities for low temperatures and large densities. [The opacities
(OPAL 95 combined with opacities of Dave Alexander) have kindly been
provided by A. Weiss.] For $M< 1.5M_{\sun}$ the opacities have been slightly extrapolated.
For this reason and since complications in the equation of state (e.g.
Coulomb interactions) are ignored, the treatment of systems with small masses
is approximate only.

Figure~\ref{fig14} shows the mass ratio $q$ as a function
of the angular momentum $J_{52}$ for different masses. The upper
border of the region covered by configurations (near $q=0.9$) is determined by the
condition that the amplitude of the circulation is non-negative. In other words,
the upper border is the curve $f=0$ which is invariant.
For smaller values of the temperature difference
the upper border occurs at larger mass ratios.

The lower border is the line $q=0.1$ since the Chebyshev approximation
for the neck used in the numerical code (c.f. K1) is valid only for $q\ge 0.1$.
An exception concerns small masses ($M\le 1.3 M_{\sun}$). For these masses
the lower border, occuring at somewhat larger mass ratios,
is caused by the prescribed temperature difference. Below the border the
difference is bound to be larger than 200~K.

The relation between angular momentum and mass ratio for given mass is unique.
Since the derivative ${\rm d}q/{\rm d}J_{52}$ is positive,
the system with the largest mass ratio has the largest angular momentum.
Systems of sufficiently large angular momentum (or mass ratio) are stable,
while the remaining systems are unstable.

When passing over from intermediate to small masses the border between
stable and unstable systems is shifting towards large mass ratios.
For a mass somewhat smaller than the solar mass stability becomes impossible.
This holds as well for smaller values of $\Delta T_{\rm e}$.
Accordingly, there is a lower limit for the mass of unevolved contact binaries,
and thus also a lower limit for the effective temperature. The limit is
caused not by the lack of configurations but by their instability.

Figure~\ref{fig15} shows the period-mass ratio diagram. Solid lines represent again
systems of positive charge.  They do not cross, which implies that the structure
of an unevolved system in thermal equilibrium
with $Z=0.02$ is (in a reasonable approximation) fully
determined by two observable properties. In the mass range under consideration
there is a minimum period of about 0.18 days
which is compatible with the observed minimum of 0.221 days for CC Com. The
maximum period  (about 0.48 days) however is far below the observed maximum.

Figures~\ref{fig16}--\ref{fig18} show characteristic properties in dependence on
mass and mass ratio. The results for the degree of contact $F$ offer an
observational test. Observed values for
$F$ are usually small compared to unity.
This is in accordance with the results in Fig.~\ref{fig18} since for all stable systems
$F$ is small compared to unity. This lends support for the treatment of the stability
problem.
Recall that a rather large value for the efficiency ($f_{\rm E}=0.1$) is used.
As shown in Table~\ref{obs1} a reduced efficiency leads to an enhanced (but still
small) degree of contact.

\begin{figure}
\psscalefirst
\psfig{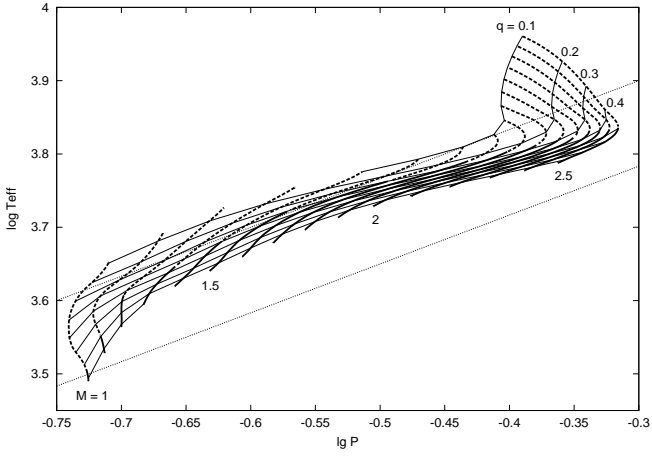}
\caption[]{The period-colour relation for unevolved systems with $Z=0.02$.
Shown are lines of constant mass (strong lines, solid/dashed for
positive/negative charge) and of constant mass ration (weak lines).
Observed systems occur in the strip limited by the dotted lines.}
\label{fig22}
\end{figure}

\begin{figure}
\psscalefirst
\psfig{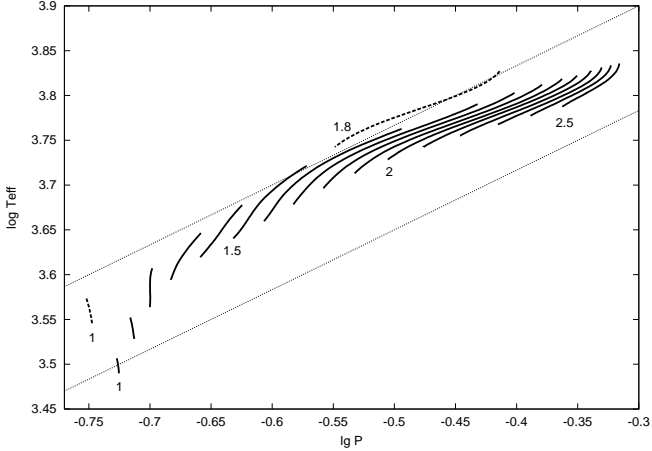}
\caption[]{The period-colour relation for stable unevolved systems with $Z=0.02$
(solid lines) and $Z=0.01$ (dashed lines).
The numbers denote the mass in solar units.}
\label{fig23}
\end{figure}

\begin{figure}
\psscalefirst
\psfig{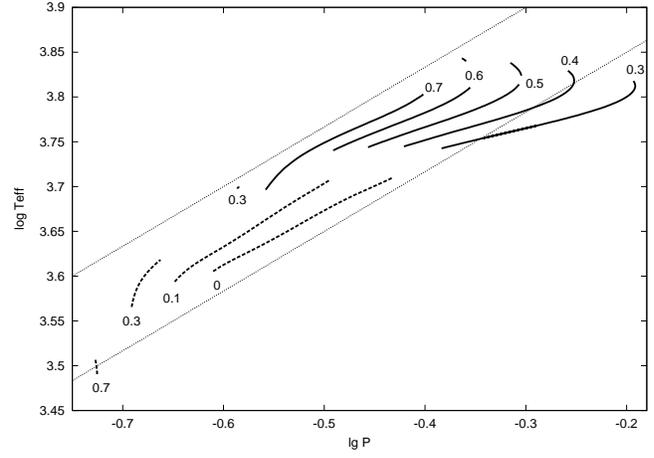}
\caption[]{The period-colour relation for stable evolved systems with $Z=0.02$ and 
the masses $M=1M_{\sun}$
(dashed lines) and $M=1.8M_{\sun}$ (solid lines).
The numbers denote the hydrogen content in the primary's centre.}
\label{fig24}
\end{figure}

%%%%%%%%%%%%%%%%%%%%%%%%%%%%%%%%%%%%%%%%%%%%%%%%%%%%%%%%%%%%%%%%%%%%%%%%%%%%%%%

\subsection{Effects of metallicity changes}  \label{effectmetal}

The results obtained so far do not cover the range of
observed systems. This can best be seen in the period-mass ratio diagram (Fig.~\ref{fig15}).
Systems having large periods and intermediate or small mass ratios are absent.
Stable systems with very small mass ratios are also absent. Since systems with these
properties are observed, effects neglected so far have to be taken into account.
Here we consider the effects of metallicity changes in unevolved systems.

The effects of a decrease in $Z$  turned out
to be small, apart from an increase in luminosity and
effective temperature for given mass and mass ratio. Figure~\ref{fig19}
shows examples in
in the period-mass ratio diagram. Results for $M=1M_{\sun}$
represent very small masses where the difficulties are largest. Results for
$M=1.8M_{\sun}$ represent typical masses.  In both cases the influence of
$Z$ is weak only. In particular, apparently the lower limit for the period
depends only slightly on $Z$. The absence of stable systems with large periods and
intermediate or small mass ratios is even more pronounced when $Z=0.01$.
We conclude that metallicity effects cannot help to extend the range of
stable systems in the period-mass ratio diagram.

%%%%%%%%%%%%%%%%%%%%%%%%%%%%%%%%%%%%%%%%%%%%%%%%%%%%%%%%%%%%%%%%%%%%%%%%%%%%%%

\subsection{Effects of evolution}  \label{effectevol}

Next we investigate evolutionary effects as the only remaining possibility
to extend the range of stable systems. In the component $i$ the simple hydrogen profile
\begin{equation}                            \label{prof}
X=\left\{ \begin{array}{ll}
X_{{\rm c}i}+(0.7-X_{{\rm c}i})\,\sin\left[ \pi x_i/(2x_{\rm c})\right]&
\quad\mbox{if \, $x_i\le x_{\rm c}$}\\
 0.7                                             & \quad \mbox{otherwise} \\
              \end{array}
     \right.
\end{equation}
will be assumed, depending on the central hydrogen content $X_{{\rm c}i}$
and a parameter $x_{\rm c}$. The standard value $x_{\rm c}=0.5$ will be used.
Effects of changes  in $x_{\rm c}$ will be investigated
when analysing individual observed systems in Sect~\ref{test}.

In the general case both components are evolved and the evolutionary status
is described by two parameters $X_{{\rm c}1},X_{{\rm c}2}$.
In an approximate treatment evolutionary effects in the secondary can be neglected
unless the mass ratio is large, and the evolutionary status is described by
one parameter $X_{{\rm c}1}$.

Concerning the choice of the circulation function we checked the invariance
and found that the use of the standard circulation function is justified also in
evolved systems.

Evolutionary effects for systems with $M=1M_{\sun},Z=0.02$ are shown in Fig.~\ref{fig20}.
Except for the thin solid line all curves have been calculated neglecting
evolutionary effects in the secondary.
Heavy lines represent configurations with a given value for $X_{{\rm c}1}$.
As before, solid/dashed lines represent configurations of positive/negative
charge. Evolved configurations with very large mass ratios ($q\ga 0.8$) have
been omitted.
With decreasing $X_{{\rm c}1}$ the curves are shifting to the right, i.e.
to larger periods, and the extent of stable configurations is increasing. Stable
evolved systems with very small mass ratios exist for periods larger than
about 0.3 days.

On the dotted curves the angular momentum is constant.
Since in the course of evolution $X_{{\rm c}1}$ decreases and the angular momentum 
does not increase, the period increases and the mass ratio decreases.
The lower limit for the period found in unevolved systems is therefore valid also
for evolved systems.

In systems with large mass ratios the results are rough only since
effects of nuclear evolution in the secondary are neglected. The influence of
these effects can be seen if they are overestimated. This is the case
in the thin solid line, which represents stable configurations with
$X_{{\rm c}1}=X_{{\rm c}2}=0.3$ and $q\la 0.8$. The curves of constant
angular momentum for $X_{{\rm c}1}=X_{{\rm c}2}$ almost coincide with the curves
for $X_{{\rm c}2}=0.7$. Accordingly, the main results obtained  neglecting
evolutionary effects in the secondary (increase of period and decrease of mass
ratio in the course of evolution) remain valid if these
effects are taken into account.

A similar diagram for systems with $M=1.8M_{\sun}$ is shown in Fig.~\ref{fig21}.
Again it can be seen that the evolution (with or without loss of angular
momentum) proceeds towards larger periods and smaller mass ratios.
Again stable evolved systems with very small mass ratios are found.
Most important is the result that large periods and small mass ratios are possible 
among evolved systems.

We conclude that the structure of contact binaries depends sensitively on
evolutionary effects and that the range of observed systems
can be explained only by these effects.

%%%%%%%%%%%%%%%%%%%%%%%%%%%%%%%%%%%%%%%%%%%%%%%%%%%%%%%%%%%%%%%%%%%%%%%%%%%%%%

\subsection{The period-colour relation}  \label{per}

The period-colour relation of observed contact binaries
(Eggen \cite{eggen1},\cite{eggen7}) provides an important observational
test. In the period-colour diagram (Figs.~\ref{fig22}--\ref{fig24})
the observed systems are mainly found in the strip
limited by the dotted lines. Figure~\ref{fig22} shows stable and unstable unevolved
configurations with $Z=0.02$. Almost all configurations outside the strip
are unstable.

Let the strip be divided in two strips of the same width (upper and lower strip,
respectively). Figure~\ref{fig23} shows that stable unevolved systems cover roughly
the upper strip, and Fig.~\ref{fig24} shows that stable evolved systems cover
almost the full strip. Accordingly, the full strip is covered by stable systems,
and stable systems outside the strip are almost absent.

These results lend support to the assumption of thermal equilibrium as well as
to the treatment of the stability problem.

%%%%%%%%%%%%%%%%%%%%%%%%%%%%%%%%%%%%%%%%%%%%%%%%%%%%%%%%%%%%%%%%%%%%

\begin{table}
\caption[ ]{Models for AB And.} \label{AB}
\begin{flushleft}
\begin{tabular}{llllllll}
\hline
\noalign{\smallskip}
 $f_{\rm E}$ & $\Delta T_{\rm e}$ & $Z$ & $T_{\rm e}$ & $L$ & $R_1$ & $R_2$ & $F$ \\
\noalign{\smallskip}
\hline
\noalign{\smallskip}
          &     &       & 5700 & 1.46 & 1.05  & 0.76  & 0.15 \\
$10^{-1}$ & 200 & 0.010 & 5756 & 1.57 & 1.022 & 0.735 & 0.037 \\
$10^{-2}$ & 200 & 0.010 & 5760 & 1.58 & 1.025 & 0.738 & 0.061 \\
$10^{-3}$ & 200 & 0.010 & 5760 & 1.60 & 1.031 & 0.744 & 0.099 \\
$10^{-4}$ & 200 & 0.010 & 5759 & 1.64 & 1.040 & 0.754 & 0.162 \\
$10^{-4}$ & 200 & 0.011 & 5714 & 1.59 & 1.040 & 0.754 & 0.160 \\
$10^{-4}$ & 100 & 0.011 & 5734 & 1.63 & 1.046 & 0.760 & 0.198 \\
\noalign{\smallskip}
\hline
\end{tabular}
\end{flushleft}
\end{table}

\begin{table}
\caption[ ]{Models for BV Dra.} \label{BV}
\begin{flushleft}
\begin{tabular}{llllllll}
\hline
\noalign{\smallskip}
 $f_{\rm E}$ & $\Delta T_{\rm e}$ & $Z$ & $T_{\rm e}$ & $L$ & $R_1$ & $R_2$ & $F$ \\
\noalign{\smallskip}
\hline
\noalign{\smallskip}
          &     &       & 6190 & 2.29 & 1.12  & 0.76  & 0.114 \\
$10^{-1}$ & 200 & 0.004 & 6333 & 2.49 & 1.091 & 0.723 & 0.050 \\
$10^{-1}$ & 200 & 0.005 & 6251 & 2.36 & 1.091 & 0.723 & 0.048 \\
$10^{-1}$ & 200 & 0.006 & 6172 & 2.24 & 1.091 & 0.723 & 0.047 \\
$10^{-1}$ & 200 & 0.007 & 6095 & 2.13 & 1.090 & 0.723 & 0.046 \\
$10^{-2}$ & 200 & 0.006 & 6175 & 2.27 & 1.095 & 0.728 & 0.079 \\
$10^{-3}$ & 200 & 0.006 & 6174 & 2.30 & 1.102 & 0.735 & 0.129 \\
$10^{-3}$ & 100 & 0.006 & 6192 & 2.35 & 1.106 & 0.739 & 0.158 \\
\noalign{\smallskip}
\hline
\end{tabular}
\end{flushleft}
\end{table}

\begin{table}
\caption[ ]{Models for BW Dra.} \label{BW}
\begin{flushleft}
\begin{tabular}{llllllll}
\hline
\noalign{\smallskip}
 $f_{\rm E}$ & $\Delta T_{\rm e}$ & $Z$ & $T_{\rm e}$ & $L$ & $R_1$ & $R_2$ & $F$ \\
\noalign{\smallskip}
\hline
\noalign{\smallskip}
          &     &       & 5930 & 1.40 & 0.98  & 0.55  & 0.140 \\
$10^{-1}$ & 200 & 0.006 & 5900 & 1.33 & 0.965 & 0.537 & 0.053 \\
$10^{-2}$ & 200 & 0.006 & 5903 & 1.35 & 0.968 & 0.541 & 0.088 \\
$10^{-3}$ & 200 & 0.006 & 5904 & 1.37 & 0.973 & 0.546 & 0.144 \\
$10^{-3}$ & 200 & 0.005 & 5978 & 1.44 & 0.974 & 0.547 & 0.146 \\
$10^{-3}$ & 100 & 0.005 & 5995 & 1.46 & 0.977 & 0.550 & 0.181 \\
\noalign{\smallskip}
\hline
\end{tabular}
\end{flushleft}
\end{table}

\begin{table*}
\caption[ ]{Models for OO Aql.} \label{OO}
\begin{flushleft}
\begin{tabular}{lllllllll}
\hline
\noalign{\smallskip}
 $f_{\rm E}$ & $Z$ & $X_{{\rm c}2}$ & $X_{{\rm c}1}$ &  $T_{{\rm e}1}$ & $L$ & $R_1$ & $R_2$ & $F$ \\
\noalign{\smallskip}
\hline
\noalign{\smallskip}
          &       &       &       & 5700 & 3.30 & 1.39  & 1.29  & 0.27  \\
$10^{-2}$ & 0.010 & 0.185 & 0.170 & 5939 & 3.54 & 1.318 & 1.218 & 0.061 \\
$10^{-2}$ & 0.012 & 0.170 & 0.157 & 5836 & 3.30 & 1.318 & 1.218 & 0.060 \\
$10^{-3}$ & 0.012 & 0.160 & 0.147 & 5827 & 3.33 & 1.328 & 1.228 & 0.097 \\
$10^{-4}$ & 0.013 & 0.137 & 0.124 & 5758 & 3.26 & 1.345 & 1.245 & 0.159 \\
$10^{-5}$ & 0.013 & 0.100 & 0.103 & 5736 & 3.36 & 1.375 & 1.275 & 0.264 \\
$10^{-5}$ & 0.014 & 0.095 & 0.096 & 5685 & 3.24 & 1.373 & 1.274 & 0.259 \\
\noalign{\smallskip}
\hline
\end{tabular}
\end{flushleft}
\end{table*}

%%%%%%%%%%%%%%%%%%%%%%%%%%%%%%%%%%%%%%%%%%%%%%%%%%%%%%%%%%%%%%%%%%%%%%%%%%%%%%%%%%%%%%

\section{Individual observational tests}
\label{test}

In this section individual observed systems with reliable data will be used as
tests of the theory. First we shall ask whether such tests are severe and what we can
learn from them.

\subsection{Observable quantities and degrees of freedom}   \label{quant}

We begin with an observational difficulty.
A light curve synthesis requires an assumption concerning the distribution of
surface brightness. Usually a gravity brightening law is assumed
(sometimes with different exponents in the two components)
and the temperature difference is treated as an adjustable parameter.
(In particular, a negative temperature difference is adopted
to reproduce the light curves of W-type systems.)
The arbitrariness of this procedure is manifest. If a gravity brightening
law exists there is no freedom in the temperature difference.

Actually there is no sound theoretical argument in favour of a correlation
between local gravity and local effective temperature.
In a rotating single star without complications (e.g. spots)
an (individual) gravity brightening law exists
already for symmetry reasons. In a contact binary symmetry arguments fail
as well as physical arguments (c.f. K\"ahler \& Fehlberg \cite{kf91}),
and a detailed mapping of the surface appears to be necessary as recognized
already by Hilditch et al. (\cite{hil}).

In the present treatment the temperature difference is bound to be positive,
and there is no freedom to explain the light curves of W-type systems
with a negative difference. A negative difference is also in conflict with
infrared observations (e.g. Hrivnak (\cite{hri89}) and UV observations
(e.g. Ruci\'{n}ski \cite{ru93}).

In view of these difficulties we consider the observed mean temperature
as  more reliable than the temperature difference, and the
observed total luminosity $L$ as more reliable than the luminosities
$L_1,L_2$ of the components.

Period $P$, mass ratio $q$, mass $M$, mean temperature $T_{\rm e}$, luminosity $L$,
radii $R_i$, and degree of contact $F$ will be taken as observational quantities.
They are not independent. Recall that $R_i=\lambda_i A$, where the separation
$A$ of the components is a function of $M$ and $P$ on account of Kepler's law.
The normalized radius $\lambda_i$
is a function of $q$ and $F$. The luminosity $L$ is known if radii and temperature are
known. We are left with $N_{\rm obs}=5$ independent observables $P,q,M,T_{\rm e},F$.

The degrees of freedom in a model are mass $M$, angular momentum
(or period $P$), metallicity $Z$, central hydrogen content in the primary $X_{{\rm c}1}$,
and efficiency $f_{\rm E}$. (There is little freedom in the temperature difference,
and the effects of changes in $\alpha$ are unimportant). Accordingly, there are
$N_{\rm free}=5$ degrees of freedom.

Since $N_{\rm free}\ge N_{\rm obs}$ there is freedom to
fit a  model to the observations. The tests involving individual systems
are therefore not able to provide evidence for thermal equilibrium.
Nevertheless they are important. Since $N_{\rm free}=N_{\rm obs}$
the model is well-determined. The structure of observed
systems can therefore be determined, assuming thermal equilibrium,
and the stability can be tested.
In the case of instability either the treatment of the
stability problem is insufficient or the system is in thermal imbalance.
In the case of stability the efficiency can be calibrated.

\begin{figure}
\psscalefirst
\psfig{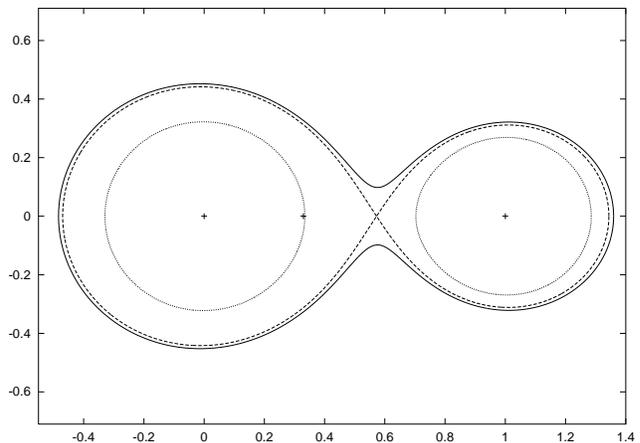}
\caption[]{The structure of AB And in the equatorial plane. Shown are the surface
(solid line), the critical surface (dashed line), and the lower borders of the
convective envelopes (dotted lines). The dots denote the centres of the components
and the centre of mass.}
\label{fig25}
\end{figure}

\subsection{The system AB And}            \label{ab}
For the system AB~And ($P=0.331892$ day) Hrivnak (\cite{hri})
determined the mass ratio $q=0.491$, the mass $M=1.5M_{\sun}$, and other
properties listed in the first line of Table~\ref{AB}, where
temperatures are in Kelvin and luminosities and radii are in solar units.
(This holds as well in the following tables.) The temperature is an
average between the components.
According to Hrivnak the uncertainty in $L$ is large.
Standard errors in the last digits are
1 for the radii and 3 for the degree of contact.

Treating the observed values for period, mass, and mass ratio as
constraints and using $X_{{\rm c}1}$ to adjust the mass ratio,
a configuration is determined by
the metallicity $Z$ and the parameters $f_{\rm E},\Delta T_{\rm e}$.
Models for AB And are summarized in Table~\ref{AB}. The charge is positive and
the models are indeed stable. This has been checked by evolutionary calculations
following a small perturbations. The amplitude of the circulation is small
($f\simeq 0.04$).
The models show that the theory is compatible with the
observations. The results for the  temperature suggests the metallicity  $Z=0.011$.
The results for the degree of contact suggest a very small value for
the efficiency ($f_{\rm E}\simeq 10^{-4}$). The model with these two
values and with $\Delta T_{\rm e}=200$~K
is in excellent agreement with the observations, except for the radii
which are somewhat too small. We shall return to this point.

The geometry of the system in the equatorial plane is shown in Fig.~\ref{fig25}.
Note that the fractional extent in radius of the convective envelope is
much larger in the primary than in the secondary. The fractional extent in mass
is also much larger.

We checked also the effects of changes in the chemical profile.
In the models listed in Table~\ref{AB}, calculated with the standard profile 
($x_{\rm c}=0.5$), the hydrogen content $X_{{\rm c}1}$ is somewhat lower than 0.4.
Models with $x_{\rm c}=0.3$ give similar results with $X_{{\rm c}1}\simeq 0.2$.
Accordingly, the choice of the chemical profile is of minor importance.
The amount of hydrogen already converted to helium is also similar
and can be used to estimate the age of the system.

\subsection{The system BV Dra}            \label{bv}
The contact binaries BV Dra and BW Dra are particularly interesting since
they form a visual binary. Precise observational results have been
obtained by Kaluzny \& Rucinski (\cite{karu}). We begin with BV~Dra.
Period (0.350067 day), mass ($1.47M_{\sun}$) and mass ratio ($q=0.411$)
are again treated as constraints. Other observational results are listed
in the first line of Table~\ref{BV}.
Standard errors in the last digits are 21 for the luminosity, 1 for the radii,
and 27 for the degree of contact.

The models of BV turned again out to be stable. Comparing the temperatures with
the observed value we obtain the metallicity $Z=0.006$, in accordance with the
result of Kaluzny \& Rucinski. The degree of contact suggests again a very
low efficiency ($f_{\rm E}\simeq 10^{-3}$). The model with these values and
with $\Delta T_{\rm e}=200$~K is in very good agreement with the observations,
again apart from the fact that the radii are somewhat too small.

\begin{figure}
\psscalefirst
\psfig{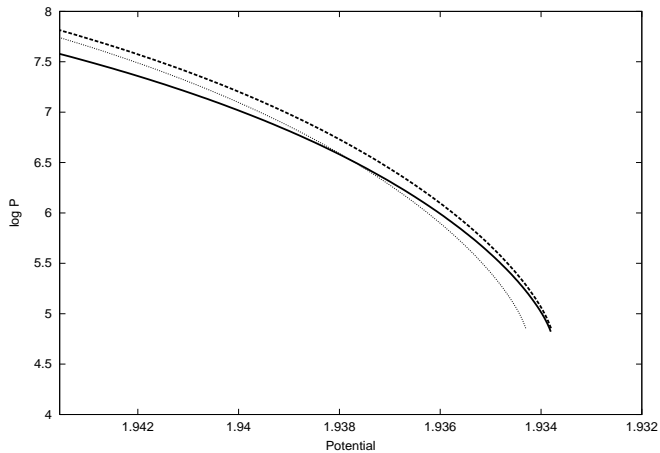}
\caption[]{$\log P$ as a fuction of the potential in the common envelope of
system 2. The strong solid/dashed line represents the primary/secondary.
The dotted line represents a modified run of pressure in the secondary (see text).}
\label{fig26}
\end{figure}

\subsection{The system BW Dra}            \label{bw}
The constraints for this system are $P=0.292167$ day, $M=1.18 M_{\sun}$,
$q=0.280$. Other observational results are listed in the first line of
Table~\ref{BW}. Standard errors in the last digits are 1 for the radii,
15 for the luminosity, and 34 for the degree of contact.
Since no mean temperature has been reported by Kaluzny \& Rucinski,
a temperature somewhat smaller than the primary's polar temperature
($5980$~K) has been adopted, as it is the case in BV Dra.

The models are again stable. This has been checked in evolutionary calculations.
The temperature suggests a metallicity
$Z=0.005\ldots 0.006$ which is very similar to the metallicity of BV Dra.
This is as expected. The degree of contact suggests again
the efficiency $f_{\rm E}\simeq 10^{-3}$. Adopting these values
and a temperature difference $\Delta T_{\rm e}=200$~K,
the agreement between theory and observations is very close.

\subsection{The system OO Aql}        \label{oo}
The system OO Aql, observed by Hrivnak~(\cite{hri89}), has a very large mass
ratio ($q=0.843$). Other constraints are $M=1.92 M_{\sun}$, $P=0.506789$ day.
The observed temperature difference $\Delta T_{\rm e}=65$~K
has been added as a constraint. Other observed properties
are listed in the first line of Table~\ref{OO}.

Since $q$ is large, evolutionary effects in the secondary are important.
(If they are ignored, the constraints require a negative amplitude $f$.)
Table~\ref{OO} contains properties of configurations
for different combinations of $X_{{\rm c}1}$ and $X_{{\rm c}2}$.
These combinations are approximately determined by the condition that $f$
is positive and small.
Note that $X_{{\rm c}2}$ is only slightly larger than $X_{{\rm c}1}$
and that these values depend on the choice of $x_{\rm c}$.
Evolutionary effects are large in this system, in accordance with the result
of Hrivnak. The configurations in the last two lines are in excellent
agreement with the observations. The radii are slightly smaller than
those given by Hrivnak. The agreement is nevertheless perfect.
Hrivnak's radii are volume radii (c.f. Mochnacki \cite{moch}) and thus larger
(in the present case by about 1.4 percent) than
the radii in a spherically averaged treatment of the components
(c.f. K\"ahler \cite{k86}). (A similar difference in radii occurs in the
systems discussed above). The metallicity ($Z=0.013\ldots0.014$) turns out
to be larger than estimated by Hrivnak. The efficiency is extremely small
($f_{\rm E}\simeq 10^{-5}$), and the amplitude is rather large ($f\simeq 0.09$).

Again the stability was tested in evolutionary calculations. The two best
models (in the last two lines of Table~\ref{OO}) are stable,
but the other models are unstable. Apparently an extremely low efficiency
(or a rather large degree of contact)
is necessary for stability. This is suggested also by other models not
listed in Table~\ref{OO}.

\subsection{Conclusions}                     \label{conclu}

Several observed systems with well-determined parameters
have been used as tests of the theory.
In all cases an excellent agreement between theory and
observations is obtained. This is as expected from the number of
free parameters. In all systems evolutionary effects are important.

All systems turned out to be stable. This lends support
not only to the transport equation used in the present model but also to the
assumption of thermal equilibrium and to the treatment of the stability problem.

The efficiency was determined to  be very small
($f_{\rm E}\simeq 10^{-3}\ldots 10^{-5}$).
This shows that the velocities of the internal mass motions in the
neighbourhood of the inner Lagrangian point are much smaller than the
sound velocity.

%%%%%%%%%%%%%%%%%%%%%%%%%%%%%%%%%%%%%%%%%%%%%%%%%%%%%%%%%%%%%%%%%%%

\begin{figure}
\psscalefirst
\psfig{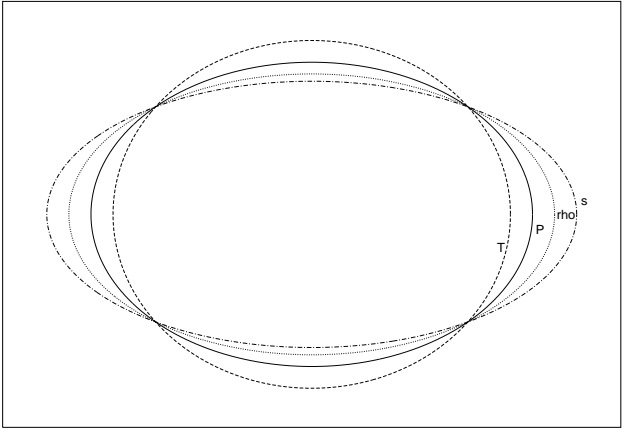}
\caption[]{Surfaces of constant pressure (solid line), constant temperature (dashed line),
constant entropy (dash-dotted line), and constant density (dotted line)
in a radiative region of a rotating star.}
\label{fig27}
\end{figure}

\section{Towards an understanding of the internal mass motions}
\label{towa}

\subsection{Internal mass motions in the common envelope} \label{com}

So far Roche geometry has been assumed and the effects of internal mass motions
on the equation of hydrostatic equilibrium have been neglected. This is
justified as a close approximation to real systems, but a closer look
into the layers in the common envelope reveals an inconsistency.
This inconsisteny can be removed by a slight modification, and this
procedure offers insight into the internal mass motions in the
common envelope.

As noted in Sect.~\ref{int}, in the present approximate treatment the
pressure difference $\Delta P$ on equipotential surfaces is negative apart from the
surface. For the layers in the common envelope of
system~2 this is shown by the strong lines in Fig.~\ref{fig26}.
Since the large-scale mass motions are driven by pressure differences
on equipotential surfaces, and since this difference is negative in each layer,
mass is flowing from the secondary to the primary but not from the primary to the 
secondary. This is
incompatible with a configuration in thermal equilibrium with
a vanishing net rate of mass transfer.

Can this inconsistency be removed by a slight modification of the structure 
equations, taking into account internal mass motions? 
The pressure difference in the common envelope is largest (apart from the sign)
in the layers just above the critical surface, causing a mass flow from the 
secondary to the primary in these layers. In a slightly changed and more
consistent treatment this feature is preserved. We need therefore
a flow from the primary to the secondary in the outermost layers
in order to obtain a vanishing net rate. The resulting velocity field
is of the type discussed by Nariai~(\cite{nar}) and Zhou \& Leung  (\cite{zhle}).

The effects of this velocity field on the run of pressure in the common envelope
are determined by a modified surface condition (the generalization of the
Roche equipotential condition), a change of the effective gravity, and (since the 
motions are expected to be turbulent) the presence of a turbulent pressure.
We shall discuss these effects in turn. Note that the Roche potential is
well defined and that the difference between Roche potential and effective potential
is negligible in the common envelope. The pressure can therefore be treated as a 
function of the potential. Since we ask for slight changes, we may assume that
the change of a curve in Fig.~\ref{fig26} is a superposition of a horizontal shift
(caused by the change in the surface condition) and a change in the pressure
gradient (caused by the change in the effective gravity and by the turbulent
pressure). To reduce the inconsistency we need a shift to the left/right
or a decrease/increase in the gradient of the secondary's/primary's curve.

Since the velocity field has a reversing layer, from Bernoulli's equation
the surface condition
\begin{equation} \label{surf}
\Psi_1+\frac{\langle v_1^2\rangle}{2}=\Psi_2+\frac{\langle v_2^2\rangle}{2}
\end{equation}
is obtained, where $v_i$ is the velocity at the surface of the component~$i$ and the angle
brackets denote a spherical average, cf. K\"ahler (\cite{k95}). Since the 
pressure vanishes at the surfaces, the surface condition acts as a boundary 
condition for the run of pressure. Assuming a horizontal shift of the curves,
it can be veryfied that a reduction of the inconsistency requires that
\begin{equation}              \label{hori}
\langle v_2^2\rangle>\langle v_1^2\rangle,
\end{equation}
i.e. that the velocities in the secondary are larger than in the primary.
The dotted line in Fig.~\ref{fig26} shows an example for the result of a
horizontal shift of the secondary's curve. It is manifest that
the inconcistency can be removed in this way.

Consider next the effective gravity. In the presence of internal mass motions
$g_i$ has to be replaced by
\begin{equation}               \label{effg}
\tilde{g}_i=\frac{{\rm G}m_i}{r_i^2}-\frac{2}{3}\Omega_i^2 r_i,\quad
{\rm with} \quad\Omega_i^2>\omega^2,
\end{equation}
i.e. the effective gravity is reduced (K\"ahler \cite{k95}). This implies that
also the pressure gradient is reduced. If this occurs in the secondary,
again the inconsistency is reduced since the pressure at the critical surface
decreases.

It remains to discuss the turbulent pressure. Let $P_{{\rm g},i}$ be the 
gas pressure and $P_i=f_iP_{{\rm g},i}$ the total pressure. Neglecting the 
radiation pressure we have $f_i=1$ in the absence of internal mass motions 
and $f_i\simeq 2$ if the turbulent velocities are comparable with the sound 
velocity. Combining the equation of hydrostatic equilibrium and the equation 
of state we find
\begin{equation}         \label{comb}
\frac{\rm d}{{\rm d}r_i}\ln P_i=
-\frac{\mu_i}{\Re T_i} \frac{\tilde{g}_i}{f_i}.
\end{equation}
Accordingly, an increase in $f_i$ has the same effect as a decrease in the
effective gravity. Turbulent pressure reduces the pressure gradient. If this 
occurs mainly in the secondary, the inconsistency is again reduced.

The discussion of the three effects has shown that the inconsistency can be 
removed by mass motions which are larger in the secondary than in the primary.
Pressure differences in the layers
avove the critical surface provide the driving mechanism for the velocity
field in the common envelope.  The result is a field with a reversing
layer, with motions from the secondary to the primary in the region above the
critical surface and from the primary to the secondary in the surface layers,
and with velocities which are larger in the secondary than in the primary.
Martin \& Davey (\cite{marda}) obtained in hydrodynamic simulations a velocity
field with the same properties.

The arguments presented in this subsection concern
late-type as well as early-type systems. They are not conclusive since
in a spherically averaged treatment of the components degrees of freedom
of the internal mass motions escape notice.

\begin{figure}
\psscalefirst
\psfig{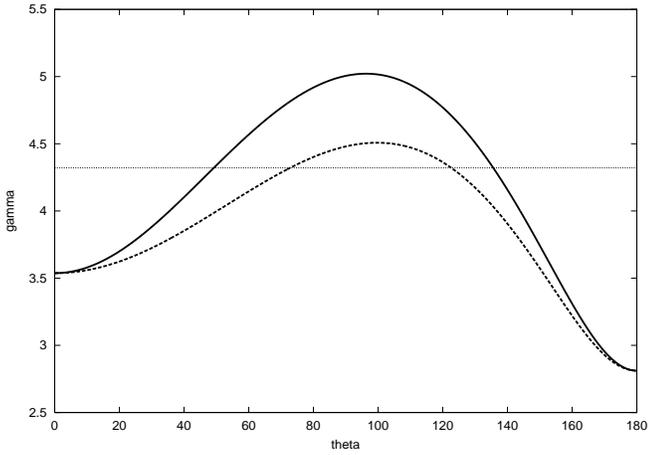}
\caption[]{The normalized gravity $\gamma$ at the top of the radiative interior
in the secondary of AB And, in dependence of the angular distance (in degrees) 
$\vartheta$ from the $x$-axis.
Shown is the run of $\gamma$ in the plane containing the rotation axis (solid line)
and in the equatorial plane (dashed line). The dotted line shows the averaged
gravity.}
\label{fig28}
\end{figure}

\subsection{Circulation currents in single stars} \label{sec}

Consider first the radiative envelope of an early-type single star in almost
uniform rotation. Near the boundary to the convective core as well as near the surface
viscous effects are important as discussed by Tassoul \& Tassoul (\cite{tass95}),
but in the bulk of the envelope the original inviscid solution of Sweet (\cite{sweet})
is a reasonable approximation. Accordingly, on an isobaric surface the temperature is
higher around the  poles (where the gravity is larger) than around the equator
(where the gravity is smaller). The resulting topology of the surfaces of constant
thermodynamic quantities is shown in Fig.~\ref{fig27}. The eccentricity is smallest
in the surface of constant temperature and largest in the surface of constant entropy.
We shall need the result that also on a surface
of constant entropy the temperature is larger around the poles than around the
equator.

Let $l(s)$ be the luminosity carried through the surface $s={\rm const}$, which
is the sum
\begin{equation}    \label{sums}
l(s)=l_{\rm rad}(s)+l_{\rm circ}(s)
\end{equation}
of the radiative luminosity  $l_{\rm rad}$ and
the  circulation luminosity $l_{\rm circ}$. In thermal equilibrium and outside the
core with nuclear burning we have ${\rm d}l/{\rm d}s=0$ and thus
\begin{equation}    \label{abl}
{\rm d}l_{\rm circ}/{\rm d}s=-{\rm d}l_{\rm rad}/{\rm d}s.
\end{equation}
Assuming stationary currents the heat equation reduces to
\begin{equation}    \label{heat}
{\rm div}{\bf F}=-\rho T {\bf u} \cdot {\rm grad}\,s,
\end{equation}
where ${\bf F}$ denotes the radiative flux and ${\bf u}$ is the circulation velocity.
We integrate this equation over the volume enclosed by the surfaces $s=s_1$ and
$s=s_2$, where the difference $s_2-s_1$ is positive and infinitesimal. The result is
\begin{equation}    \label{resu}
l_{\rm rad}(s_2)-l_{\rm rad}(s_1)=
-(s_2-s_1)\int_{s=s_2}\rho T\, ({\bf u}\cdot {\bf n})\,{\rm d}A,
\end{equation}
where ${\bf n}={\rm grad}\, s/|{\rm grad}\, s|$ is the outward directed normal on the
surface $s=s_2$ and ${\rm d}A$ is an element of this surface. Let
$\bar{T}$ be the average of $T$ on this surface and $\Delta T=T-\bar{T}$.
Since the net mass flow trough the surface vanishes we obtain
\begin{equation}    \label{arri}
  \frac{{\rm d}l_{\rm rad}}{{\rm d}s}
=-\int \Delta T \rho\, ({\bf u}\cdot {\bf n})\,{\rm d}A.
\end{equation}
Since $\Delta T$ and ${\bf u}\cdot {\bf n}$ have the same sign almost everywhere,
positive around the pole and negative around the equator,
the integrand is positive almost everywhere.
Therefore ${\rm d}l_{\rm rad}/{\rm d}s$ is negative, and from Eq.~(\ref{abl})
we see that ${\rm d}l_{\rm circ}/{\rm d}s$ is positive. A slightly generalized
discussion shows that this result remains valid in the presence of nuclear burning.

We conclude that radiative layers are sources of the circulation luminosity.
The circulation luminosity is positive and  a monotonically
increasing function of $s$ or of the mass variable.
These results concern the bulk of the radiative envelope where viscosity
is unimportant, but not the regions at the borders of the envelope. Recall that
sinks near the surface are needed.

\subsection{Circulation currents in contact binaries} \label{curr}
First let us ask for the changes of the gravity ${\bf g}$ on a level surface in the
secondary's interior. The mass $m_2$ enclosed by this surface is known from the
spherically averaged treatment of the system. In a close approximation
${\bf g}$ is (apart from the sign) the gradient of the potential
\begin{equation}            \label{pot}
-\frac{{\rm G}M_1}{d_1}-\frac{{\rm G}m_2}{d_2}-\frac{1}{2}\omega^2 d^2,
\end{equation}
where $d_i$ is the distance from the centre of the component $i$ and $d$ is the distance
from the rotation axis. 

Taking the secondary of AB And as an example,
Fig.~\ref{fig28} shows the normalized gravity $\gamma=gA^2/({\rm G}M)$
at the  bottom of the convective envelope,
i.e. at the top of the radiative interior.
The dashed line shows $\gamma$ in the equatorial plane in dependence on the 
angular distance $\vartheta$ from the $x$-axis. (As usual, the $x$-axis is taken to lie
along the line of centres of the components, with the origin in the primary's centre.)
The solid line shows the run of $\gamma$ in
the plane containing the rotation axis, and the dotted line
denotes the averaged gravity obtained in a spherical treatment of the
components. The curves show that the gravity changes almost by a factor of two.
On level surfaces deep in the interior the variation is smaller, but on the surface
having (in a spherical treatment) the fractional radius 0.5  the gravity still changes
by about 10 percent.

On each level surface the gravity is largest at the poles (solid curve for
$\vartheta=90^\circ$),
small near $\vartheta=0^\circ$, and smallest when $\vartheta=180^\circ$,
i.e. when closest to the inner Lagrangian point. On account of tidal synchronization
the rotation is known to be almost uniform. Although
the departures from sphericity in the outer layers are large and although
the geometry is more complex than in single stars, the essential features are similar.
We therefore expect rising motions at the poles and sinking
motions at the equator, at least near the $x$-axis. Similar motions
are expected also in the primary.

This is supported by results on radiative regions in uniform
rotation (Mestel \cite{mest}). They predict
a double-cell pattern with a circulation inversion on the level surface with the
density $\rho=\omega^2/(2\pi{\rm G})$. In AB And the outer cell is absent in the
primary (since the critical density occurs in the convective envelope) and
unimportant in the secondary since the bulk of the radiative interior is covered by
the inner cell. In this cell the circulation is upward/downward in regions
of higher/lower gravity. Mestel's result concerns an idealized problem
since barotropy is assumed and viscosity is neglected. Questions about
the validity of Mestel's result (Tassoul \& Tassoul \cite{tass95})
concern the outer cell and the region around the circulation inversion,
but not the bulk of the radiative region where Sweet's (\cite{sweet}) results
are confirmed.

Accordingly, it is certain that on on an isobaric surface the density is lower
in regions of outward motions than in regions of inward motions.
Therefore the discussion of the preceding subsection applies, showing
that there are sources of the circulation luminosity throughout the bulk of
the radiative interior. This is in line with the basic assumption
made in Sect.\ref{intro}. It is also in line with the
result from Sect.~\ref{inv} that stable systems have sources
not only in the outer layers but also deep in the interior.
The corresponding sinks occur either
in the outermost radiative layers or in the common turbulent envelope.

%%%%%%%%%%%%%%%%%%%%%%%%%%%%%%%%%%%%%%%%%%%%%%%%%%%%%%%%%%%%%%%%%%%%%%

\section{Summarizing and concluding remarks}  \label{concl}

The present investigation was based on the assumption that the circulation
luminosity in the secondary's interior is positive and large enough to enable thermal
equilibrium. The first part of this assumption can be abandoned.
As shown in the preceding section the circulation luminosity is certainly positive
in the radiative interior, not only in the secondary but also in
the primary. In systems with intermediate or small mass ratios the effects
in the primary are small while the
effects in the secondary are important since the core luminosity is small.
In systems with large mass ratios (e.g. OO~Aql) the neglect of the circulation
luminosity in the primary introduces a slight error.

The resulting treatment of contact binaries is simple. Sources of the circulation
luminosity deep in the secondary's interior are necessary, but details are
unimportant. Free parameters are the temperature difference between the components
$\Delta T_{\rm e}$, bound to be positive and small,
and the efficiency $f_{\rm E}$ of the energy transfer between the components.
The fractional extent of radiative regions is larger in the secondary than in
the primary. In the course of evolution the period increases and the mass ratio 
decreases.

Adopting standard values for the parameters, a survey of unevolved and evolved
contact configuration has been obtained. The results are
successful in explaining the basic observational facts, in particular the preference
for shallow contact, the minimum period and the minimum temperature, the existence of
stable configurations with mass ratios very close to unity, and the period-colour
relation. Since stability considerations are essential in these observational tests
we consider the results as strong arguments not only in favour of thermal equilibrium
but also in favour of the treatment of the stability problem.

Assuming thermal equilibrium, models of individual observed systems with reliable
 data are well-determined
apart from some freedom in $\Delta T_{\rm e}$. When stable they can be used to
to calibrate the efficiency and to determine metallicity and age.
The models obtained so far are stable,
which lends again support to the assumption of thermal equilibrium.
Evolutionary effects are important. The efficiency is very small
($f_{\rm E}=10^{-3}\ldots 10^{-5}$).

Concerning the hydrodynamic problem, arguments have been presented that the
velocity field in the common envelope has a reversing layer with motions from
the secondary to the primary in the layers just above the critical surface and
from the primary to the secondary in the surface layers.

Details of the circulation, in particular the circulation function, can
be determined only in a three-dimensional discussion. In comparison with rotating
stars the problem is more difficult since the geometry is more complex.
In other respects the problem is simpler since rotation law and geometry
are known in a close approximation. This problem will be investigated in a 
forthcoming paper, again assuming thermal equilibrium.

Although we have strong arguments in favour of thermal equilibrium we have no proof, and
it is uncertain whether a proof is possible.

\listofobjects
\end{document}